\documentclass{IEEEtran}
 \pdfoutput=1
\usepackage[T1]{fontenc}
\usepackage{color}
\usepackage{float}
\usepackage{mathtools}
\usepackage{amsmath}
\usepackage{amsthm}
\usepackage{amssymb}
\usepackage{stackrel}
\usepackage{graphicx}
\usepackage[unicode=true,
 bookmarks=true,bookmarksnumbered=true,bookmarksopen=true,bookmarksopenlevel=1,
 breaklinks=false,pdfborder={0 0 0},pdfborderstyle={},backref=false,colorlinks=false]
 {hyperref}
\hypersetup{pdftitle={Your Title},
 pdfauthor={Your Name},
 pdfpagelayout=OneColumn, pdfnewwindow=true, pdfstartview=XYZ, plainpages=false}

\makeatletter

\providecommand{\tabularnewline}{\\}
\floatstyle{ruled}
\newfloat{algorithm}{tbp}{loa}
\providecommand{\algorithmname}{Algorithm}
\floatname{algorithm}{\protect\algorithmname}

\theoremstyle{plain}
\newtheorem{thm}{\protect\theoremname}

\usepackage[caption=false,font=normalsize,labelfont=sf,textfont=sf]{subfig}

\makeatother

\providecommand{\theoremname}{Theorem}

\begin{document}
\title{Multi-model Stochastic Particle-based Variational Bayesian Inference
for Multiband Delay Estimation}
\author{Zhixiang~Hu,~An~Liu,~\IEEEmembership{Senior Member,~IEEE}, and
Minjian Zhao,~\IEEEmembership{Member,~IEEE}\thanks{Zhixiang Hu, An Liu, and Minjian Zhao are with the College of Information
Science and Electronic Engineering, Zhejiang University, Hangzhou
310027, China, e-mail: \protect\href{http://anliu@zju.edu.cn}{anliu@zju.edu.cn}.}}
\maketitle
\begin{abstract}
Joint utilization of multiple discrete frequency bands can enhance
the accuracy of delay estimation. Although some unique challenges
of multiband fusion, such as phase distortion, oscillation phenomena,
and high-dimensional search, have been partially addressed, further
challenges remain. Specifically, under conditions of low signal-to-noise
ratio (SNR), insufficient data, and closely spaced delay paths, accurately
determining the model order\textemdash the number of delay paths\textemdash becomes
difficult. Misestimating the model order can significantly degrade
the estimation performance of traditional methods. To address joint
model selection and parameter estimation under such harsh conditions,
we propose a multi-model stochastic particle-based variational Bayesian
inference (MM-SPVBI) framework, capable of exploring multiple high-dimensional
parameter spaces. Initially, we split potential overlapping primary
delay paths based on coarse estimates, generating several parallel
candidate models. Then, an auto-focusing sampling strategy is employed
to quickly identify the optimal model. Additionally, we introduce
a hybrid posterior approximation to improve the original single-model
SPVBI, ensuring overall complexity does not increase significantly
with parallelism. Simulations demonstrate that our algorithm offers
substantial advantages over existing methods.
\end{abstract}


\section{Introduction}

\IEEEPARstart{H}{igh-accuracy} sensing, as one of the key new abilities
in next-generation multi-functional wireless networks, will dramatically
enrich the range of services that wireless networks can provide, such
as WiFi sensing \cite{MB_ESPRIT}, vehicular networks, integrated
sensing and communications (ISAC) \cite{ISAC}, internet of things
(IoT) \cite{IOT}, among others. For range-based localization, it
is vital to achieve high-accuracy estimates of the multipath channel
delays. Generally speaking, a larger signal bandwidth tends to enhance
the precision of delay estimation. However, the finite spectral resources
and hardware constraints restrict the signal bandwidth from being
excessively large. To further improve the accuracy of delay estimation,
multiband delay estimation has been recently proposed \cite{MB_ESPRIT,MB_VTM,wyb_MB,SPVBI_JIOT,MB_add2}.
This technique, by leveraging measurements of channel state information
(CSI) across multiple non-contiguous bands, offers the potential to
boost resolution and range accuracy.

To fully exploit the multiband benefits, we must tackle some unique
challenges. Firstly, there is a phase misalignment and timing synchronization
error across different frequency bands. Secondly, the multi-dimensional
parameters make the search space of the estimation problem huge. Thirdly,
high-frequency carriers induce a severe oscillation in the likelihood
function, resulting in numerous suboptimal local peaks in proximity
to the true delay value \cite{MB_VTM}. In \cite{SPVBI_JIOT}, these
issues have been substantially mitigated through a two-stage Bayesian
approach and a particle-based posterior approximation method.

However, a tricky challenge persists before the multi-path delay estimation
\textendash{} the necessity to determine the dimension of the parameter
vector of the data model, i.e., the model order \cite{ModelSel_book}.
Given the inherent multipath conditions and the complexities of radio
propagation, signals may reach the receiver through various uncertain
paths, in addition to the line-of-sight (LoS) path. Therefore, we
need to first extract the number of unknown delay paths from the data.

Conventionally, to estimate the model's order, we employ certain selection
criteria \cite{Info_criterion,Info_criterion2}, such as the Akaike
information criterion (AIC) \cite{AIC1974,AIC1998}, minimum description
length (MDL) criterion \cite{MDL,MB_MDL_AIC} or Bayesian information
criterion (BIC) \cite{BIC}. These information criteria typically
consist of two terms: one item that aims to maximize likelihood, and
another that imposes a penalty on model complexity. Their intrinsic
simplicity has led to their widespread usage. Then, parameter estimation
is performed based on this determined order. In most instances, this
sequential process proves to be feasible.

However, if the SNR is low, observational data is insufficient, or
there are some overlapping delay paths, the model order may not be
accurately determined, since these information criteria are asymptotically
effective. Additionally, some prior information about the number of
multipaths cannot be fully utilized. Therefore, to enhance estimation
accuracy, joint model selection and parameter estimation are considered.

In terms of joint model selection and parameter estimation, existing
algorithms mainly fall into two categories: one is based on Markov
chain Monte Carlo (MCMC) sampling \cite{MCMC}, while the other is
based on support vector.

(1) A typical MCMC-based algorithm is the reversible jump MCMC (RJMCMC)
algorithm \cite{RJMCMC1995,RJMCMC1999}. RJMCMC algorithm facilitates
Bayesian inference by dynamically exploring multiple model spaces,
offering a capacity to alter model complexity via reversible jumps
between parameter spaces. While the prior information about the number
of multi-path can be exploited, this algorithm still exhibits certain
shortcomings. RJMCMC requires numerous iterations (often in the tens
of thousands) to converge to the steady state of the Markov chain,
and bears a high computational complexity, demanding a large number
of samples to approximate the posterior distribution. The subsequent
population Monte Carlo (PMC) algorithm \cite{PMC,S-PMC} reduced iteration
counts but still faces challenges in high-dimensional sampling and
proposal distribution selection, particularly for complex likelihood
functions.

(2) Another category of methods is based on support vector. This approach
\cite{wyb_MB,support_VBI} employs a sparse linear representation
framework, where signal parameters are discretized into an observation
matrix and sparse support vectors. Through iterative optimization,
the sparsity of support vectors inherently identifies multipath components,
enabling delay estimation via the dictionary structure. The advantage
of such algorithms is that they do not require extensive sampling;
instead, the posterior estimates of the model order and target parameters
are directly obtained through alternate iterations. However, there
are certain drawbacks: the model order has an upper limit constraint,
and the algorithm, being based on the expectation-maximum (EM) algorithm
framework, can easily fall into local optimum. For instances of delay
overlapping and non-linear models, support vector-based algorithms
can not work well.

In this paper, for the issue of joint model selection and parameter
estimation in the multiband delay estimation scenarios, a multi-model
stochastic particle-based variational Bayesian inference (SPVBI) algorithm
is proposed. The proposed multi-model SPVBI (MM-SPVBI) can be viewed
as an extension of the SPVBI algorithm \cite{SPVBI_JIOT}, advancing
it from a single-model framework to a multi-model one.

The MM-SPVBI maintains multiple latent models in parallel and gradually
focuses on the parameter space with the maximum posterior. The objective
is to resolve potential overlapping delay paths, and to achieve super-resolution
posterior estimations even under low SNRs. The MM-SPVBI addresses
key limitations of existing methods: (1) compared to support vector-based
approaches, it employs VBI to effectively handle nonlinear models
while preventing local optima through two-stage estimation; (2) compared
to RJMCMC, it achieves comparable estimation accuracy with substantially
reduced computational complexity. In addition, compared to the original
SPVBI, MM-SPVBI has improved sampling methods and posterior approximation
techniques, generating a low-complexity algorithm suitable for high-dimensional
and non-convex likelihood functions with multiple peaks. The main
contributions are summarized below.
\begin{itemize}
\item \textbf{A multi-model stochastic particle-based VBI method. }Building
upon the SPVBI algorithm, MM-SPVBI algorithm has been proposed to
tackle the issue of inaccurate model order under low SNRs in multiband
delay estimation. The MM-SPVBI algorithm showcases high flexibility
and efficiency by integrating techniques such as variational inference,
particle approximation, and stochastic successive convex approximation
(SSCA). As such, it is suited for multiband delay estimation problems
that are high-dimensional and non-convex. Furthermore, the joint optimization
of model order and parameters could enable the algorithm to achieve
higher estimation accuracy and resolve overlapped delay paths.
\item \textbf{Auto-focusing sampling strategy for stochastic optimization.}
We introduce stochastic successive convex approximation to address
the tricky crux of multiple integrals in Bayesian inference. In this
process, the gradient of the objective function is approximated through
stochastic sampling. However, in the case of multiple model Bayesian
inference, the sampling overhead is considerable, which necessitates
the adoption of a new sampling strategy. Samples are drawn from parameter
spaces with varying dimensions, and the number of samples allocated
dynamically based on the posterior distributions of different models.
This approach allows computational resources to be focused on high-probability
parameter spaces, thereby enhancing the efficiency of sampling approximation.
\item \textbf{Hybrid posterior approximation.} If all variables are characterized
by a particle-based discrete distribution as in \cite{SPVBI_JIOT},
the complexity will dramatically increase when transitioning from
a single-model space to multi-model parameter spaces. Therefore, on
the basis of the all-particle posterior approximation scheme, we propose
a hybrid approach. Here, some variables of multi-peak distributions
are approximated by particles, while others are represented by Gaussian
distributions. The comprehensive use of diverse posterior approximation
schemes can improve the fitting efficiency and reduce the computational
complexity in MM-SPVBI.
\end{itemize}
The rest of the paper is organized as follows. The system model and
problem formulation are given in Section $\text{\mbox{II}}$. The
multi-model framework is detailed in Section $\text{\mbox{III}}$,
while the MM-SPVBI algorithm implementation and analysis are presented
in Section $\text{\mbox{IV}}$. Numerical simulations and performance
validation are provided in Section $\text{\mbox{V}}$. Finally, conclusions
are given in Section $\text{\mbox{V}I}$.

Notations: $\propto$ denotes the left is proportional to the right,
$\delta(\cdot)$ denotes the Dirac\textquoteright s delta function,
$vec\left[\cdot\right]$ denotes the vectorization operation, and
$\left\Vert \cdot\right\Vert $ denotes the Euclidean norm. $\mathbf{A}^{T}$,
$\mathbf{A}^{H}$, $\mathbf{A}^{-1}$ represent a transpose, complex
conjugate transpose, and inverse of a matrix $\mathbf{A}$. $\mathbb{E}_{z}[\cdot]$
denotes the expectation operator with respect to the random vector
$z$. $\boldsymbol{D}_{KL}\left[q||p\right]$ denotes the Kullback-Leibler
(KL) divergence of the probability distributions $q$ and $p$. $\mathcal{N}(\mu,\Sigma)$
and $\mathcal{CN}(\mu,\Sigma)$ denotes Gaussian and complex Gaussian
distribution with mean $\mu$ and covariance matrix $\Sigma$.

\section{System Model and Problem Formulation\label{sec:Problem-Formulations}}

\subsection{System Model}

\begin{figure}[t]
\begin{centering}
\textsf{\includegraphics[scale=0.35]{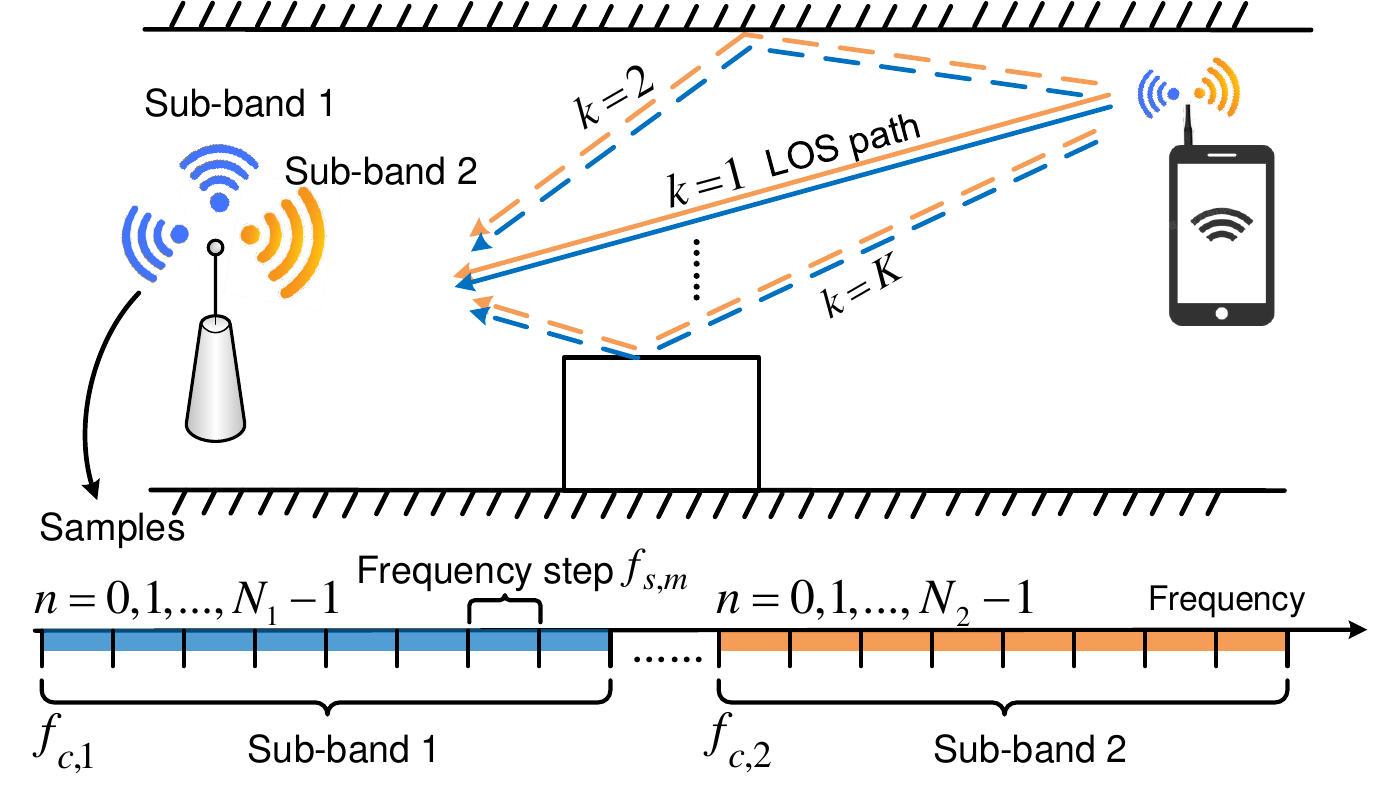}}
\par\end{centering}
\caption{\textsf{\label{MB}}Illustration of multiband sensing and spectrum
distribution.}
\end{figure}

In a multiband delay estimation scenario, a single-input single-output
(SISO) system is considered in Fig. \ref{MB}, which utilizes known
OFDM training signals across $M$ frequency subbands to estimate the
range between the mobile node and receiver. After processing, the
discrete received signal in frequency domain can be expressed as \cite{SPVBI_JIOT,MB_VTM}
\begin{equation}
\begin{split}y_{m}^{\left(n\right)} & =\sum\limits_{k=1}^{K}\alpha_{k}e^{-j2\pi\left(f_{c,m}+nf_{s,m}\right)\tau_{k}}e^{-j2\pi nf_{s,m}\delta_{m}}e^{j\phi_{m}}+w_{m}^{\left(n\right)},\end{split}
\label{eq:original_model}
\end{equation}
where $m=1,2,\ldots,M$ is the frequency band index, $N_{m}$ denotes
the number of subcarriers in the $m$-th subband. $n=0,1,\ldots,N_{m}-1$
is the subcarrier index. $k=1,2,\ldots,K$ denotes the $k$-th scattering
path. $\alpha_{k}$ is the complex gain that carries the amplitude
and phase information of a scattering path \footnote{In our narrowband scenario (bandwidth $\ll$ coherence bandwidth),
the frequency-flat channel assumption holds, making the frequency-dependent
complex gain variation negligible \cite{dependence}.}, and $\tau_{k}$ is the time delay of the $k$-th path. $f_{c,m}$
and $f_{s,m}$ denote the initial frequency and subcarrier spacing
of the $m$-th subband, respectively. Additionally, hardware imperfections
introduce a random initial phase $\phi_{m}$ (caused by oscillator
variations \cite{wyb_MB}) and a timing synchronization error $\delta_{m}$
(resulting from packet detection delay and sampling frequency offset
\cite{wyb_MB,nonidealfoctor3}) into CSI measurements. $w_{m}^{\left(n\right)}$
represents additive white Gaussian noise (AWGN) following $\mathcal{CN}\left(0,\sigma_{w}^{2}\right)$.

Following the two-stage delay estimation scheme in \cite{SPVBI_JIOT},
we can split the target parameter estimation into two stages, each
utilizing distinct signal models derived from the original model in
\eqref{eq:original_model}. In the coarse estimation stage, by absorbing
the random phase $e^{j\phi_{m}}$ and carrier phase $e^{-j2\pi f_{c,m}\left(\tau_{k}+\delta_{m}\right)}$
of each subband into the complex gain $\alpha_{k}$, the original
signal model can be reformulated as the following coarse estimation
signal model:
\begin{equation}
\begin{split}y_{m}^{\left(n\right)} & =\sum\limits_{k=1}^{K}\alpha_{k,m}^{'}e^{-j2\pi(nf_{s,m})\left(\tau_{k}+\delta_{m}\right)}+w_{m}^{\left(n\right)},\end{split}
\label{eq:coarse_model}
\end{equation}
where $\alpha_{k,m}^{'}=\alpha_{k}\left(e^{j\phi_{m}}e^{-j2\pi f_{c,m}\tau_{k}}\right)$.
The \textquotedblleft term absorption\textquotedblright{} refers to
the process of consolidating specific components within the model
into new composite terms, which are then estimated and processed as
unified entities. This merging operation enables the remaining parts
of the signal model to exhibit distinct functional characteristics.
In the coarse estimation model, this absorption creates a quasi-linear
structure that facilitates the application of conventional estimation
algorithms. More significantly, compared to the GHz-level carrier
frequencies $\left(f_{c,m}+nf_{s,m}\right)$, the MHz-level bandwidth
aperture $\left(nf_{s,m}\right)$ becomes substantially smaller. Consequently,
the likelihood function for delay no longer exhibits severe oscillations,
which effectively eliminates numerous undesirable local optima and
significantly narrows the subsequent search space \cite{SPVBI_JIOT}.

In the refined estimation stage, taking the first frequency band as
the reference band, we absorb the random phase $e^{j\phi_{1}}$ and
carrier phase $e^{-j2\pi f_{c,1}\tau_{k}}$ of the reference band
into $\alpha_{k}$. This results in the following refined signal model:
\begin{equation}
\begin{split}y_{m}^{\left(n\right)} & =\sum\limits_{k=1}^{K}\alpha_{k}^{'}e^{-j2\pi\left(f_{c,m}'+nf_{s,m}\right)\tau_{k}}e^{j\phi_{m}^{'}}e^{-j2\pi nf_{s,m}\delta_{m}}+w_{m}^{\left(n\right)},\end{split}
\label{eq:refined_model}
\end{equation}
where $\alpha_{k}^{'}=\alpha_{k}\left(e^{j\phi_{1}}e^{-j2\pi f_{c,1}\tau_{k}}\right)$,
$\phi_{m}^{'}=\phi{}_{m}-\phi_{1}$, $f_{c,m}'=f_{c,m}-f_{c,1}$,
$\phi_{1}^{'}=0$, $f_{c,1}^{'}=0$. The preservation of phase differences
serves to eliminate potential phase ambiguity between $\phi{}_{m}$
and $\alpha_{k}$, as detailed in Section III-A of \cite{wyb_MB}.
The construction of frequency-difference term $\left(f_{c,m}'+nf_{s,m}\right)$
serves to sharpen the likelihood function for delay estimation while
preventing excessive oscillations.

By equipping different signal models, the two-stage scheme can narrow
the parameter range during the coarse estimation stage, and avoiding
local optima during the refined estimation stage. Consequently, we
can jointly leverage the carrier phase information across different
frequency bands (i.e., multiband gain) to enhance the accuracy of
delay estimation.

\subsection{Problem Formulation}

The coarse estimation stage employs well-established weighted root
MUSIC (WR-MUSIC) and LS methods to obtain preliminary parameter estimates
(including $\left|\hat{\alpha}_{k}\right|,\hat{\tau}_{k},\hat{\phi}'_{m},\hat{\delta}_{m}$)
along with their estimation intervals \cite{SPVBI_JIOT}. This established
approach allows our research to focus primarily on developing advanced
algorithms in the refined estimation stage that specifically address
the critical challenge of delay path aliasing resolution. In the refined
signal model \eqref{eq:refined_model}, the variables to be estimated
are denoted as $\mathbf{\boldsymbol{\theta}}{\rm =}\left[\alpha_{1}^{'},\ldots,\alpha_{K}^{'},\tau_{1},\ldots,\tau_{K},\phi_{2}^{'},\ldots,\phi_{M}^{'},\right.$
$\left.\delta_{1},\ldots,\delta_{M}\right]^{T}$, and $J=\left|\mathbf{\boldsymbol{\theta}}\right|$
is the number of variables, which varies with the model order (i.e.,
the value of $K$). Without loss of generality, we assume that $\tau_{1}<\tau_{2}<\ldots<\tau_{K}$.
Therefore, $\tau_{1}$ represents the delay of the LoS path, which
is essential for ranging.

However, achieving high precision in estimating $\tau_{1}$ necessitates
an accurate knowledge of the model order before the coarse estimation.
Under challenging conditions such as low SNR, insufficient observation
data, or presence of overlapping delay paths, the model order determination
may become unreliable. When the number of fitted multipath components
in the received signal deviates from the true value, the estimation
of the target parameter is likely to deviate significantly from its
true value. Typically, when $K$ is unknown, we employ classical criteria
such as AIC, MDL, and BIC to first determine the model order $K$.
These criteria are generally formulated as follows \cite{Info_criterion,Info_criterion2}:
\begin{equation}
\hat{K}=\textrm{arg}\,\underset{K}{\textrm{min}}-2\,\textrm{ln}p\left(\boldsymbol{y}|\hat{\mathbf{\boldsymbol{\theta}}},\boldsymbol{\wp}_{v}\right)+\eta\cdot K,\label{eq:criteria}
\end{equation}
where $\textrm{ln}p\left(\boldsymbol{y}|\hat{\mathbf{\boldsymbol{\theta}}},\boldsymbol{\wp}\right)$
represents the log-likelihood function of the given model $\boldsymbol{\wp}\in\mathcal{M}$
with the maximum likelihood (ML) estimate $\hat{\mathbf{\boldsymbol{\theta}}}$,
where $\mathcal{M}=\left\{ \boldsymbol{\wp}_{v},v=1,\ldots,N_{\zeta}\right\} $
is the value space, with a total of $N_{\zeta}$ candidate models.
These candidate models account for different scenarios that may arise
from delay path aliasing, where overlapping multipath components create
ambiguous channel configurations. $\wp$ quantifies the distinction
between candidate models with varying model orders (i.e., numbers
of delay paths, $K$) and diverse channel configurations under identical
or differing model orders, such as distinct combinations of $\alpha_{k}^{'}$
and $\tau_{k}$. $\mathbf{\boldsymbol{\theta}}$ is taken from the
parameter space $\boldsymbol{\Theta}$ of the given model $\boldsymbol{\wp}$.
$\boldsymbol{y}=\left[y_{1}^{\left(0\right)},y_{1}^{\left(1\right)},\ldots,y_{1}^{\left(N_{1}-1\right)},\ldots,y_{M}^{\left(0\right)},y_{M}^{\left(1\right)},\ldots,\right.$
$\left.y_{M}^{\left(N_{M}-1\right)}\right]^{T}\in\mathbb{C}^{N_{all}\times1}$
is a frequency domain observation vector, where $N_{all}=\sum_{m=1}^{M}N_{m}$.
$\eta$ is the penalty coefficient. For AIC, $\eta=2$, and for MDL/BIC,
$\eta=\textrm{ln}\left(N_{all}\right)$.

In the case of AWGN, the log-likelihood function of \eqref{eq:refined_model}
can be written as follow:
\begin{equation}
\begin{split} & \textrm{ln}p\left(\boldsymbol{y}|\mathbf{\boldsymbol{\theta}},\boldsymbol{\wp}\right)=\textrm{ln}\prod\limits_{m=1}^{M}\prod\limits_{n=0}^{N_{m}-1}p(y_{m}^{\left(n\right)}|\mathbf{\mathbf{\boldsymbol{\theta}}},\boldsymbol{\wp})\\
 & =-\sum\limits_{m=1}^{M}N_{m}\ln\left(\pi\sigma_{w}^{2}\right)-\frac{1}{\sigma_{w}^{2}}\sum_{m=1}^{M}\sum_{n=0}^{N_{m}-1}\left|y_{m}^{\left(n\right)}-s_{m}^{\left(n\right)}\left(\mathbf{\mathbf{\boldsymbol{\theta}}}\right)\right|^{2},
\end{split}
\label{eq:log_likelihood}
\end{equation}
where
\begin{equation}
\begin{split}s_{m}^{\left(n\right)}\left(\mathbf{\mathbf{\boldsymbol{\theta}}}\right) & =\sum\limits_{k=1}^{K}\alpha_{k}^{'}e^{-j2\pi(f_{c,m}'+nf_{s,m})\tau_{k}}e^{j\phi_{m}^{'}}e^{-j2\pi\left(nf_{s,m}\right)\delta_{m}}\end{split}
\label{eq:reconstruc_sig}
\end{equation}
is the received signal reconstructed from $\mathbf{\mathbf{\mathbf{\boldsymbol{\theta}}}}{\rm =}\left[\boldsymbol{\alpha}^{'T},\boldsymbol{\varLambda}^{T}\right]^{T}$.
In vector\textendash matrix form, we have
\begin{equation}
\boldsymbol{s}\left(\boldsymbol{\theta}\right)=\mathbf{D}\left(\boldsymbol{\varLambda}\right)\boldsymbol{\alpha}^{'}\in\mathbb{C}^{N_{all}\times1},\label{eq:matrix_form}
\end{equation}
where $\boldsymbol{\varLambda}{\rm =}\left[\tau_{1},\ldots,\tau_{K},\phi_{2}^{'},\ldots,\phi_{M}^{'},\delta_{1},\ldots,\delta_{M}\right]^{T}$
and $\boldsymbol{\alpha}^{'}{\rm =}\left[\alpha_{1}^{'},\ldots,\alpha_{K}^{'}\right]^{T}\in\mathbb{C}^{K\times1}$.
The matrix $\mathbf{D}\left(\boldsymbol{\varLambda}\right)$ is defined
as $\mathbf{D}\left(\boldsymbol{\varLambda}\right)=\left[d\left(\boldsymbol{\varLambda}\right)\right]_{m,n,k}\in\mathbb{C}^{N_{all}\times K}$,
where the subscripts $\left[\cdot\right]_{m,n,k}$ range over $M$,
$N_{m}$, and $K$, with each matrix element given by $d\left(\boldsymbol{\varLambda}\right)=e^{-j2\pi(f_{c,m}'+nf_{s,m})\tau_{k}}e^{j\phi_{m}^{'}}e^{-j2\pi\left(nf_{s,m}\right)\delta_{m}}$.

However, unfavorable factors such as low SNR, insufficient observational
data, and overlapping delay paths can lead to poor performance of
order criteria. Additionally, we hope to effectively utilize some
potential prior information about the model order. Therefore, we consider
using Bayesian inference for joint model selection and parameter estimation,
which involves identifying the true model $\boldsymbol{\wp}\in\mathcal{M}$
and estimating the parameters $\boldsymbol{\theta}\in\boldsymbol{\Theta}$
associated with the model.

We can regard the model order as a realization of a discrete random
variable, thus the total parameter space can be expressed as $\mathcal{M}\times\boldsymbol{\Theta}$.
It is worth noting that each $\boldsymbol{\Theta}$ may have different
dimensions and may include different parameters. For Bayesian inference
of $\boldsymbol{\theta}$ and $\boldsymbol{\wp}$, the pivotal step
is the derivation of the posterior distribution, denoted as $p\left(\boldsymbol{\theta},\boldsymbol{\wp}|\boldsymbol{y}\right)$.
By applying Bayes\textquoteright{} theorem, the posterior distribution
of the unknown parameters is formulated as $p\left(\boldsymbol{\theta},\boldsymbol{\wp}|\boldsymbol{y}\right)\propto p\left(\boldsymbol{y}|\boldsymbol{\theta},\boldsymbol{\wp}\right)p\left(\boldsymbol{\theta}|\boldsymbol{\wp}\right)p\left(\boldsymbol{\wp}\right)$,
where $p\left(\boldsymbol{\wp}\right)$ is the prior distribution
of the model, $p\left(\boldsymbol{\theta}|\boldsymbol{\wp}\right)$
is the parameter prior for a given model $\boldsymbol{\wp}$.

In the Bayesian context, the maximum a posteriori (MAP) model selection
can be expressed as: $\boldsymbol{\wp}^{*}\mathbf{=}\mathrm{arg}\mathop{\mathrm{max}}\limits_{\boldsymbol{\wp}}\int\ldots\int p\left(\boldsymbol{\theta},\boldsymbol{\wp}|\boldsymbol{y}\right)d\boldsymbol{\theta}$.
However, the main difficulty in determining the optimal model lies
in solving multi-dimensional integrals. After that, for a variable
of interest, $\theta_{j}$, the marginal posterior distribution under
the optimal model $\boldsymbol{\wp}^{*}$ can be expressed as $p\left(\theta_{j}|\boldsymbol{y},\boldsymbol{\wp}^{*}\right)\propto\int\ldots\int p\left(\boldsymbol{y}|\boldsymbol{\theta},\boldsymbol{\wp}^{*}\right)p\left(\boldsymbol{\theta}|\boldsymbol{\wp}^{*}\right)p\left(\boldsymbol{\wp}^{*}\right)d\boldsymbol{\theta}_{\sim j}$,
where $\boldsymbol{\theta}_{\sim j}$ encapsulates all other variables
except for $\theta_{j}$. Typically, deriving a closed-form solution
for $p\left(\theta_{j}|\boldsymbol{y},\boldsymbol{\wp}^{*}\right)$
is also intractable due to the complicated measurement functions and
the high-dimensional integrals over $\boldsymbol{\theta}_{\sim j}$.

Based on the above, we wish to obtain the joint MAP estimate of the
model and target parameters by solving such a maximization problem:
\begin{equation}
\begin{split}\mathcal{P}_{0}:\mathbf{\left[\boldsymbol{\theta}^{*},\boldsymbol{\wp}^{*}\right]_{MAP}=}\mathrm{arg}\mathop{\mathrm{max}}\limits_{\boldsymbol{\theta},\boldsymbol{\wp}}\: & \ln\left[p\left(\boldsymbol{y}|\boldsymbol{\theta},\boldsymbol{\wp}\right)p\left(\boldsymbol{\theta}|\boldsymbol{\wp}\right)p\left(\boldsymbol{\wp}\right)\right]\end{split}
.\label{eq:ori_MAP_problem}
\end{equation}

In the next section, we will propose a novel Bayesian algorithm to
effectively approximate the marginal posterior of the model and target
parameters.

\section{Multi-model Framework and Innovations}

The two-stage approach in \cite{SPVBI_JIOT} first estimates model
order via AIC/MDL criteria before applying coarse estimation. However,
this method struggles with closely-spaced delays under low SNR or
limited data conditions, where coarse estimation often fails to resolve
paths, leading SPVBI to converge to suboptimal solutions within an
incorrect parameter space.

To address this, we propose a multi-model SPVBI algorithm for joint
model selection and parameter estimation. This approach extends the
original single-model SPVBI by defining posterior probability distributions
over the space of possible signal structures, where the number and
values of delay paths are no longer fixed. The posterior distribution
is evaluated across a finite, disconnected union of subspaces with
varying dimensions \cite{RJMCMC1999}. We then determine the best
model and optimize parameters within these multiple high-dimensional
spaces.

Following the VBI framework outlined in SPVBI \cite{SPVBI_JIOT},
we transform posterior inference into an optimization problem in which
the independent variables are a set of flexible variational distributions
$q\left(\boldsymbol{\theta},\boldsymbol{\wp}\right)\triangleq q_{\boldsymbol{\eta}}\left(\boldsymbol{\theta}|\boldsymbol{\wp}\right)q_{\boldsymbol{\zeta}}\left(\boldsymbol{\wp}\right)$,
where $\boldsymbol{\eta}$ represents the parameters of the hybrid
variational distribution in each model, encompassing the least-squares
estimate of $\alpha_{k}^{'}$, particle positions, weights, and the
mean/variance of Gaussian distributions. Meanwhile, $\boldsymbol{\zeta}$
corresponds to the model variational distribution parameters, primarily
consisting of the weighting coefficients $\zeta_{v}$ for different
models. Their definitions can be seen in formulas \eqref{eq:eta_def}
and \eqref{eq:zeta_def}. Under the criterion of KL divergence, $q\left(\boldsymbol{\theta},\boldsymbol{\wp}\right)$
is iteratively adjusted to approximate the true posterior distribution
$p\left(\boldsymbol{\theta},\boldsymbol{\wp}|\boldsymbol{y}\right)$.
Ultimately, posterior estimates are obtained based on the converged
variational distribution $q\left(\boldsymbol{\theta},\boldsymbol{\wp}\right)$.
In the context of multi-models, KL divergence is defined as
\begin{equation}
\begin{split} & \boldsymbol{D}_{KL}\left[q||p\right]{\rm =}\int\int q_{\boldsymbol{\eta}}\left(\boldsymbol{\theta}|\boldsymbol{\wp}\right)q_{\boldsymbol{\zeta}}\left(\boldsymbol{\wp}\right)\textrm{ln}\frac{q_{\boldsymbol{\eta}}\left(\boldsymbol{\theta}|\boldsymbol{\wp}\right)q_{\boldsymbol{\zeta}}\left(\boldsymbol{\wp}\right)}{p\left(\boldsymbol{\theta},\boldsymbol{\wp}|\boldsymbol{y}\right)}d\boldsymbol{\theta}d\boldsymbol{\wp}\\
 & {\rm =}\int\int q_{\boldsymbol{\eta}}\left(\boldsymbol{\theta}|\boldsymbol{\wp}\right)q_{\boldsymbol{\zeta}}\left(\boldsymbol{\wp}\right)\textrm{ln}\frac{q_{\boldsymbol{\eta}}\left(\boldsymbol{\theta}|\boldsymbol{\wp}\right)q_{\boldsymbol{\zeta}}\left(\boldsymbol{\wp}\right)p\left(\boldsymbol{y}\right)}{p\left(\boldsymbol{y}|\boldsymbol{\theta},\boldsymbol{\wp}\right)p\left(\boldsymbol{\theta}|\boldsymbol{\wp}\right)p\left(\boldsymbol{\wp}\right)}d\boldsymbol{\theta}d\boldsymbol{\wp},
\end{split}
\end{equation}
where $q_{\boldsymbol{\eta}}\left(\boldsymbol{\theta}|\boldsymbol{\wp}\right)$
is assumed to be factorized as $q_{\boldsymbol{\eta}}\left(\boldsymbol{\theta}|\boldsymbol{\wp}\right)=\prod_{j=1}^{J}q_{\boldsymbol{\eta}}\left(\theta_{j}|\boldsymbol{\wp}\right)$
\cite{VBI_Beal,meanfield}. Given that $p\left(\boldsymbol{y}\right)$
is a constant independent of $q\left(\boldsymbol{\theta},\boldsymbol{\wp}\right)$,
minimizing the KL divergence is equivalent to solving the following
optimization problem:
\begin{equation}
\begin{split}\mathcal{P}_{1}:[\mathbf{\boldsymbol{\eta}}^{*},\boldsymbol{\zeta}^{*}] & \mathbf{=}\mathop{\textrm{arg}\min}\limits_{\boldsymbol{\eta},\boldsymbol{\zeta}}\boldsymbol{L}\left(\boldsymbol{\eta},\boldsymbol{\zeta}\right)\end{split}
,\label{eq:ori_VBI_problem}
\end{equation}
where
\begin{align}
 & \boldsymbol{L}\left(\boldsymbol{\eta},\boldsymbol{\zeta}\right)\nonumber \\
 & \triangleq\int\int q_{\boldsymbol{\eta}}\left(\boldsymbol{\theta}|\boldsymbol{\wp}\right)q_{\boldsymbol{\zeta}}\left(\boldsymbol{\wp}\right)\textrm{ln}\frac{q_{\boldsymbol{\eta}}\left(\boldsymbol{\theta}|\boldsymbol{\wp}\right)q_{\boldsymbol{\zeta}}\left(\boldsymbol{\wp}\right)}{p\left(\boldsymbol{y}|\boldsymbol{\theta},\boldsymbol{\wp}\right)p\left(\boldsymbol{\theta}|\boldsymbol{\wp}\right)p\left(\boldsymbol{\wp}\right)}d\boldsymbol{\theta}d\boldsymbol{\wp}.\label{eq:L_define}
\end{align}
Furthermore, we can reformulate the objective function $\boldsymbol{L}\left(\boldsymbol{\eta},\boldsymbol{\zeta}\right)$
by decoupling the model variables $\boldsymbol{\wp}$ from the estimated
parameters $\boldsymbol{\theta}$:
\begin{align}
 & \boldsymbol{L}\left(\boldsymbol{\eta},\boldsymbol{\zeta}\right)\nonumber \\
 & =\int q_{\boldsymbol{\zeta}}\left(\boldsymbol{\wp}\right)\left[\int q_{\boldsymbol{\eta}}\left(\boldsymbol{\theta}|\boldsymbol{\wp}\right)\textrm{ln}\frac{q_{\boldsymbol{\eta}}\left(\boldsymbol{\theta}|\boldsymbol{\wp}\right)}{p\left(\boldsymbol{y}|\boldsymbol{\theta},\boldsymbol{\wp}\right)p\left(\boldsymbol{\theta}|\boldsymbol{\wp}\right)}d\boldsymbol{\theta}\right]d\boldsymbol{\wp}\nonumber \\
 & +\int\left[\int q_{\boldsymbol{\eta}}\left(\boldsymbol{\theta}|\boldsymbol{\wp}\right)d\boldsymbol{\theta}\right]q_{\boldsymbol{\zeta}}\left(\boldsymbol{\wp}\right)\textrm{ln}\frac{q_{\boldsymbol{\zeta}}\left(\boldsymbol{\wp}\right)}{p\left(\boldsymbol{\wp}\right)}d\boldsymbol{\wp}\nonumber \\
 & \triangleq\int q_{\boldsymbol{\zeta}}\left(\boldsymbol{\wp}\right)\boldsymbol{L}\left(\boldsymbol{\eta}|\boldsymbol{\zeta}\right)d\boldsymbol{\wp}+\int q_{\boldsymbol{\zeta}}\left(\boldsymbol{\wp}\right)\textrm{ln}\frac{q_{\boldsymbol{\zeta}}\left(\boldsymbol{\wp}\right)}{p\left(\boldsymbol{\wp}\right)}d\boldsymbol{\wp},\label{eq:L1}
\end{align}
where $\boldsymbol{L}\left(\boldsymbol{\eta}|\boldsymbol{\zeta}\right)$
is the KL divergence under the model $\boldsymbol{\wp}$, defined
as
\begin{equation}
\boldsymbol{L}\left(\boldsymbol{\eta}|\boldsymbol{\zeta}\right)\triangleq\int q_{\boldsymbol{\eta}}\left(\boldsymbol{\theta}|\boldsymbol{\wp}\right)\textrm{ln}\frac{q_{\boldsymbol{\eta}}\left(\boldsymbol{\theta}|\boldsymbol{\wp}\right)}{p\left(\boldsymbol{y}|\boldsymbol{\theta},\boldsymbol{\wp}\right)p\left(\boldsymbol{\theta}|\boldsymbol{\wp}\right)}d\boldsymbol{\theta}.\label{eq:cond_KL}
\end{equation}

In general, due to multipath aliasing, there are only a finite number
of candidate models $\mathcal{M}=\left\{ \boldsymbol{\wp}_{v},v=1,\ldots,N_{\zeta}\right\} $.
Consequently, the variational posterior distribution $q_{\boldsymbol{\zeta}}\left(\boldsymbol{\wp}\right)$
can be represented as a discrete weighted sum:
\begin{equation}
q_{\boldsymbol{\zeta}}\left(\boldsymbol{\wp}\right){\rm =}\sum\limits_{v=1}^{N_{\zeta}}\zeta_{v}\delta\left(\boldsymbol{\wp}-\boldsymbol{\wp}_{v}\right),\boldsymbol{\wp}_{v}\in\mathcal{M},\label{eq:q_model_def}
\end{equation}
where $\zeta_{v}$ is the probability when the model variable $\boldsymbol{\wp}$
is $\boldsymbol{\wp}_{v}$, and $N_{\zeta}$ is the total number of
potential models. Therefore, the parameter of $q_{\boldsymbol{\zeta}}\left(\boldsymbol{\wp}\right)$
to be updated is
\begin{equation}
\boldsymbol{\zeta}\triangleq\left[\zeta_{1},\ldots\zeta_{v}\ldots,\zeta_{N_{\zeta}}\right]^{T}.\label{eq:zeta_def}
\end{equation}
Based on \eqref{eq:q_model_def}, we can further get
\begin{equation}
\boldsymbol{L}\left(\boldsymbol{\eta},\boldsymbol{\zeta}\right)=\sum\limits_{v=1}^{N_{\zeta}}\left[\zeta_{v}\boldsymbol{L}\left(\boldsymbol{\eta}|\boldsymbol{\zeta}_{v}\right)+\zeta_{v}\textrm{ln}\zeta_{v}-\zeta_{v}\textrm{ln}\zeta_{v}^{0}\right],\label{eq:L_disc}
\end{equation}
where model prior weights $\zeta_{v}^{0}$ are typically initialized
with equal values ($\zeta_{v}^{0}=1/N_{\zeta},\forall v$), and the
prior distribution is $p\left(\boldsymbol{\wp}\right){\rm =}\sum\limits{}_{v=1}^{N_{\zeta}}\zeta_{v}^{0}\delta\left(\boldsymbol{\wp}-\boldsymbol{\wp}_{v}\right)$.
From equation \eqref{eq:L_disc}, it can be observed that, for joint
model selection and parameter estimation, the objective function can
be regarded as a weighted sum of the conditional KL divergence under
$\boldsymbol{L}\left(\boldsymbol{\eta}|\boldsymbol{\zeta}_{v}\right)$
and the proximity to the prior distribution.

\subsection{Outline of the MM-SPVBI Algorithm Framework\label{subsec:Outline}}

\begin{figure}[t]
\begin{centering}
\textsf{\includegraphics[scale=0.52]{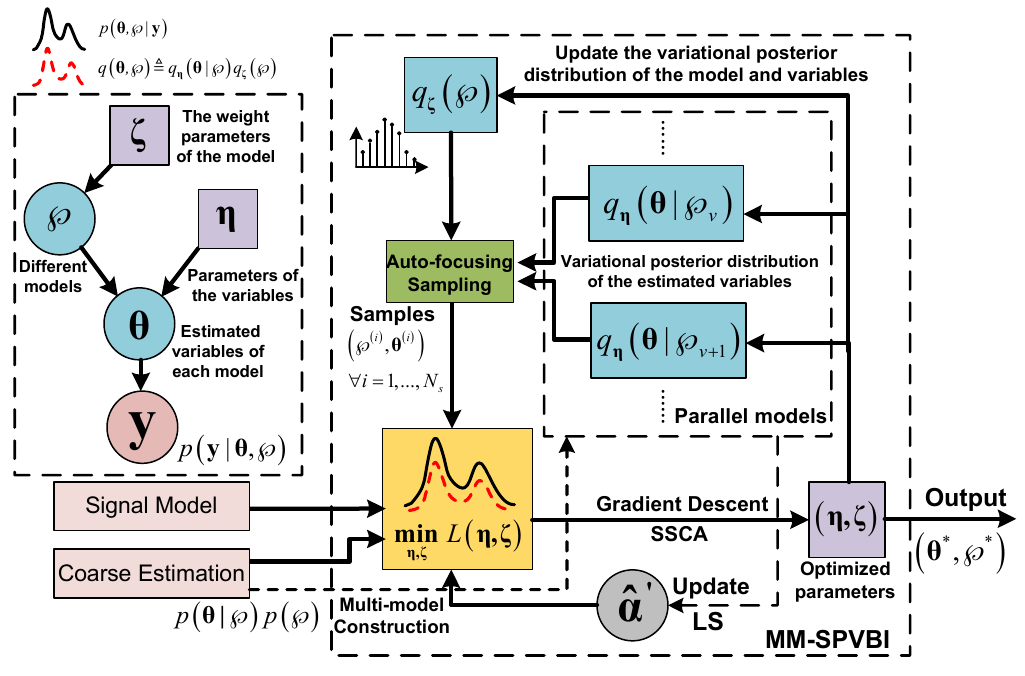}}
\par\end{centering}
\caption{\textsf{\label{MM-SPVBI}}Overall framework of the MM-SPVBI algorithm.}
\end{figure}

We first present the overall framework of the proposed MM-SPVBI algorithm,
as illustrated in Fig. \ref{MM-SPVBI}.

In MM-SPVBI algorithm, we construct multiple parallel candidate models
and initialize their parameters based on the order criteria and coarse
estimates. Each model represents a potential case of delay path aliasing,
with the parameters of individual models being updated in a similar
manner to that in single-model SPVBI. Specifically, in \cite{SPVBI_JIOT},
the SPVBI leverages particle-based discrete distributions \cite{VBI+IS}
to flexibly approximate multi-peak posterior distributions. Moreover,
by transforming the problem into a stochastic optimization task, the
block SSCA method employed \cite{SPVBI_JIOT,SSCA} can significantly
enhance sampling efficiency and reduce complexity through average-over-iteration,
making it suitable for high-dimensional problems.

Although the complexity of single-model SPVBI is not too high compared
to other multiband algorithms, searching multiple parameter spaces
still leads to a significant increase in complexity due to high parallelism.
Therefore, we have improved the full-particle posterior approximation
technique and fixed-sample-size sampling method in the SPVBI algorithm.
This improvement narrows the parameter space and facilitates rapid
pruning. Specifically, a hybrid posterior approximation and an auto-focusing
sampling strategy are employed to mitigate the complexity explosion
caused by high parallelism.

In the following subsection, we will elaborate on the detailed design
of the proposed MM-SPVBI algorithm.

\subsection{Construction of Multiple Models\label{subsec:Construction-of-Multiple}}

As illustrated in \eqref{eq:criteria}, we can initially obtain a
rough estimate of the number of delay paths using the order criterion.
However, under adverse channel conditions, this model order is prone
to be overestimated or underestimated \cite{order_estimate}.
\begin{figure*}[t]
\begin{centering}
\textsf{\includegraphics[scale=0.8]{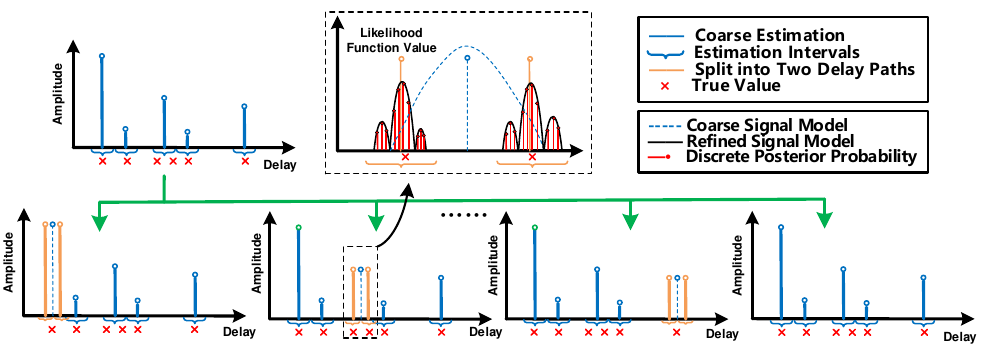}}
\par\end{centering}
\caption{\textsf{\label{Split_path}}Construct parallel models by splitting
delay paths.}
\end{figure*}

Overestimation has a limited impact, as the extra path-associated
complex gains typically converge to small values in the refined estimation
stage, albeit with increased complexity. However, underestimation
is more problematic. When the model includes fewer delay components
than are actually present, significant errors arise. If the critical
LoS path overlaps with other delay paths, the estimated composite
delay often averages the two, causing a substantial deviation in the
LoS path estimate and leading to severe performance degradation. Predicting
which delay paths will overlap is impossible, and arbitrarily selecting
more delay paths is impractical and inaccurate.

Upon determining the model order $\hat{K}$ through \eqref{eq:criteria},
preliminary estimates of the delays and complex gains can be obtained
using the WR-MUSIC algorithm and the LS method for the specified number
of paths \cite{SPVBI_JIOT}. Once the coarse estimates $\hat{\tau}_{k}$
and estimation intervals $\left[\hat{\tau}_{k}-\Delta\hat{\tau}_{k}/2,\hat{\tau}_{k}+\Delta\hat{\tau}_{k}/2\right]$\footnote{As described in \cite{SPVBI_JIOT}, the estimation interval can be
determined based on the mean squared error (MSE) of the coarse estimation
results obtained from offline training, or several times of the square
root of the Cramer-Rao lower bound (CRLB) calculated from the coarse
estimation. $\Delta\hat{\tau}_{k}$ is the length of the coarse estimation
interval for $\tau_{k}$.} of the delay paths have been identified, as shown in Fig. \ref{Split_path},
we can split the delay paths \cite{split_VB} with the highest gain
among the top candidates, treating different splitting schemes as
distinct candidate models. In general, the separation distance $\tau_{d}$
between the two split delay paths can be set as the delay resolution
of the coarse estimation algorithm, e.g., the inverse of the maximum
bandwidth of the subbands. For instance, suppose a preliminary estimation
identifies $K$ delayed paths, with their gain ranked in the order
of $\left|\hat{\alpha}_{1}^{'}\right|>\left|\hat{\alpha}_{3}^{'}\right|>\ldots$.
We can then split the top few delayed paths in terms of gain to obtain
$N_{\zeta}$ candidate models, i.e., $\left\{ \boldsymbol{\wp}_{1},\boldsymbol{\wp}_{2},\ldots,\boldsymbol{\wp}_{N_{\zeta}}\right\} \triangleq\mathcal{M}$,
where
\begin{equation}
\begin{cases}
\boldsymbol{\wp}_{1}: & \hat{\tau}_{1},\hat{\tau}_{2},\ldots,\hat{\tau}_{K}\\
\boldsymbol{\wp}_{2}: & \left(\hat{\tau}_{1}-\tau_{d}\right),\left(\hat{\tau}_{1}+\tau_{d}\right),\hat{\tau}_{2},\ldots,\hat{\tau}_{K}\\
\boldsymbol{\wp}_{3}: & \hat{\tau}_{1},\hat{\tau}_{2},\left(\hat{\tau}_{3}-\tau_{d}\right),\left(\hat{\tau}_{3}+\tau_{d}\right),\ldots,\hat{\tau}_{K}\\
 & \vdots
\end{cases}.
\end{equation}

This approach increases the likelihood of distinguishing a delay path
that might be a composite of two overlapping paths. Given the constraint
on parallelism, we generate candidate models only for the delay paths
that account for the majority of the gain, ensuring computational
complexity scales controllably with increasing numbers of multipath
components. For example, the top five delay paths in terms of gain
are each split into five candidate models. Weaker paths that might
overlap have a negligible impact on the LoS path, thus their exclusion
is reasonable. Furthermore, we only consider the case where a single
delayed path undergoes aliasing. On the one hand, if we consider the
simultaneous overlap of two or more paths, the number of models generated
by permutations becomes prohibitively large, leading to significantly
increased computational overhead. On the other hand, since the LoS
path typically exhibits higher gain, the strategy of splitting one
high-gain path already effectively captures the cases of LoS path
overlap. For time-delay estimation, the most critical element is indeed
the LoS path. Therefore, allocating more resources to identify such
overlaps and better estimate the LoS leads to significant improvement
in ranging accuracy. This emphasis is both cost-effective and efficient.

For these candidate models, we subsequently employ some hybrid variational
posterior distributions to approximate their true posterior distributions.

\subsection{Hybrid Posterior Approximation\label{subsec:Hybrid-Posterior-Approximation}}

The particle approximation in SPVBI addresses a key challenge: large
difference frequencies $(f_{c,m}'+nf_{s,m})$ create oscillatory likelihood
function spectrum for time delays $\tau_{k}$ \eqref{eq:refined_model},
complicating global optimization. As Fig. \ref{Split_path} shows,
discrete particles adaptively capture multi-peak posteriors, preventing
local optima traps. While we extend particle approximation to other
variables ($\alpha_{k}^{'}$, $\phi_{m}^{'}$, $\delta_{m}$) for
derivation consistency, their typically single-peak distributions
(unaffected by high-frequency terms) don't require full particle treatment\textemdash especially
crucial in multi-model settings where complexity scales rapidly.

For the variables $\phi_{m}^{'}$ and $\delta_{m}$, we can approximate
them using simple Gaussian distributions $\mathcal{N}\left(\mu,\sigma^{2}\right)$,
which require the update of only two parameters\textemdash the mean
$\mu$ and the variance $\sigma^{2}$\textemdash considerably fewer
than the $2N_{p}$ parameters of the particle distribution, where
$N_{p}$ is the number of particles used for each variable. Additionally,
for the complex gain $\alpha_{k}^{'}$, due to the linear structure
in \eqref{eq:matrix_form}, we can obtain a satisfactory estimate
directly by employing the LS method, obviating the need for complex
posterior inference.

Based on the analysis above, for the target variables under model
$\boldsymbol{\wp}_{v}$, we can obtain the following form of hybrid
variational posterior distribution:
\begin{align}
q_{\boldsymbol{\eta}}\left(\boldsymbol{\theta}|\boldsymbol{\wp}_{v}\right) & =\stackrel[k=1]{K}{\prod}q\left(\alpha_{k}^{'}\right)q\left(\tau_{k};\mathbf{p}_{k},\mathbf{w}_{k}\right)\nonumber \\
\cdot\stackrel[m=1]{M}{\prod} & q\left(\delta_{m};\mu_{\delta,m},\sigma_{\delta,m}^{2}\right)\stackrel[m=2]{M}{\prod}q\left(\phi_{m}^{'};\mu_{\phi,m},\sigma_{\phi,m}^{2}\right),\label{eq:hybrid_post}
\end{align}
where $q\left(\alpha_{k}^{'}\right)=\delta\left(\alpha_{k}^{'}-\hat{\alpha}_{k}^{'}\right)$
and the initial value $\hat{\alpha}_{k}^{'}$ can be obtained by LS
method after the coarse estimates $\hat{\tau}_{k}$ are determined.
$q\left(\tau_{k};\mathbf{p}_{k},\mathbf{w}_{k}\right){\rm =}\sum\limits{}_{n=1}^{N_{p}}w_{k,n}\delta\left(\tau_{k}-p_{k,n}\right)$,
$\mathbf{p}_{k}=[p_{k,1},\ldots,p_{k,N_{p}}]^{T}$ and $\mathbf{w}_{k}=[w_{k,1},\ldots,w_{k,N_{p}}]^{T}$
are the positions and weights of the particles respectively, with
particle positions $\mathbf{p}_{k}$ initialized within the estimation
intervals obtained from coarse delay estimates. $q\left(\delta_{m};\mu_{\delta,m},\sigma_{\delta,m}^{2}\right)=\mathcal{N}\left(\mu_{\delta,m},\sigma_{\delta,m}^{2}\right)$
and $q\left(\phi_{m}^{'};\mu_{\phi,m},\sigma_{\phi,m}^{2}\right)=\mathcal{N}\left(\mu_{\phi,m},\sigma_{\phi,m}^{2}\right)$,
where $\mu_{\delta,m}$ and $\sigma_{\delta,m}^{2}$ represent the
mean (set to coarse estimates $\hat{\delta}_{m}$) and variance (determined
by estimation intervals) of variable $\delta_{m}$, respectively;
while $\mu_{\phi,m}$ and $\sigma_{\phi,m}^{2}$ denote the mean (set
to coarse estimates $\hat{\phi}'_{m}$) and variance (determined by
prior knowledge) of variable $\phi_{m}^{'}$, respectively. Therefore,
the parameter $\boldsymbol{\eta}$ in each model can be denoted as
\begin{equation}
\boldsymbol{\eta}\triangleq\left[\hat{\alpha}_{1;K}^{'},\mathbf{p}_{1;K},\mathbf{w}_{1;K},\mu_{\delta,1:M},\sigma_{\delta,1:M}^{2},\mu_{\phi,2:M},\sigma_{\phi,2:M}^{2}\right]^{T},\label{eq:eta_def}
\end{equation}
where $\mathbf{p}_{1;K}$ denotes the concatenation of vectors $\mathbf{p}_{k},\forall k=1,\ldots,K$
(applying similarly to other variables).

Additionally, the prior distributions can be set as $p\left(\boldsymbol{\tau}|\boldsymbol{\wp}_{v}\right)=\prod_{k=1}^{K}\left|\Delta\tau_{k}\right|^{-1}$,
$\tau_{k}\in\left[\hat{\tau}_{k}-\Delta\hat{\tau}_{k}/2,\hat{\tau}_{k}+\Delta\hat{\tau}_{k}/2\right]$,
$p\left(\boldsymbol{\phi^{'}}|\boldsymbol{\wp}_{v}\right)$$=\left(2\pi\right)^{-\left(M-1\right)}$,
and $p\left(\boldsymbol{\delta}|\boldsymbol{\wp}_{v}\right)=\prod_{m=1}^{M}\mathcal{N}\left(0,\sigma_{0,m}^{2}\right)$
as described in \cite{SPVBI_JIOT}. Specifically, $\tau_{k}$ and
$\phi_{m}^{'}$ follow a uniform distribution, while $\delta_{m}$
follows a Gaussian distribution with small variance $\sigma_{0,m}^{2}$.
Thus, the joint prior distribution is $p\left(\boldsymbol{\theta}|\boldsymbol{\wp}_{v}\right)=p\left(\boldsymbol{\tau}|\boldsymbol{\wp}_{v}\right)p\left(\boldsymbol{\phi^{'}}|\boldsymbol{\wp}_{v}\right)p\left(\boldsymbol{\delta}|\boldsymbol{\wp}_{v}\right)$.
Based on the hybrid posterior approximation the prior distribution,
the optimization problem $\mathcal{P}_{1}$ can be reformulated as
follows:
\[
\begin{split}\mathcal{P}_{2}:\mathop{\min}\limits_{\boldsymbol{\eta},\boldsymbol{\zeta}}\quad & \boldsymbol{L}\left(\boldsymbol{\eta},\boldsymbol{\zeta}\right)=\sum\limits_{v=1}^{N_{\zeta}}\zeta_{v}\left[\boldsymbol{L}\left(\boldsymbol{\eta}|\boldsymbol{\zeta}_{v}\right)+\textrm{ln}\frac{\zeta_{v}}{\zeta_{v}^{0}}\right]\\
\text{s.t.}\quad & \sum\limits_{n=1}^{N_{p}}w_{k,n}=1,w_{k,n}\in\left[\epsilon,1\right],\quad\forall k,n,\\
 & p_{k,n}\in\left[\hat{\tau}_{k}-\Delta\hat{\tau}_{k}/2,\hat{\tau}_{k}+\Delta\hat{\tau}_{k}/2\right],\quad\forall k,n,\\
 & \mu_{\phi,m}\in\left[0,2\pi\right],\sigma_{\phi,m}^{2}\geq0,\sigma_{\delta,m}^{2}\geq0\quad\forall m,\\
 & \sum\limits_{v=1}^{N_{\zeta}}\zeta_{v}=1,\quad\zeta_{v}\in\left[\epsilon,1\right],\quad\forall v,
\end{split}
\]
where the weights $w_{k,n}$ and $\zeta_{v}$ are subject to simplex
constraints, while the particle positions $p_{k,n}$ are confined
within a coarse estimation interval $\left[\hat{\tau}_{k}-\Delta\hat{\tau}_{k}/2,\hat{\tau}_{k}+\Delta\hat{\tau}_{k}/2\right]$.
The mean $\mu_{\phi,m}$ of the random phase $\phi_{m}^{'}$ may be
restricted within the range of $0$ to $2\pi$, and the variances
$\sigma_{\phi,m}^{2}/\sigma_{\delta,m}^{2}$ are non-negative.

For the integral term within the objective function $\boldsymbol{L}\left(\boldsymbol{\eta},\boldsymbol{\zeta}\right)$,
specifically defined as the conditional KL divergence $\boldsymbol{L}\left(\boldsymbol{\eta}|\boldsymbol{\zeta}_{v}\right)$
in \eqref{eq:cond_KL}, it can be regarded as the objective function
of a stochastic optimization problem:
\begin{align}
\boldsymbol{L}\left(\boldsymbol{\eta}|\boldsymbol{\zeta}_{v}\right) & \triangleq\int q_{\boldsymbol{\eta}}\left(\boldsymbol{\theta}|\boldsymbol{\wp}_{v}\right)\textrm{ln}\frac{q_{\boldsymbol{\eta}}\left(\boldsymbol{\theta}|\boldsymbol{\wp}_{v}\right)}{p\left(\boldsymbol{y}|\boldsymbol{\theta},\boldsymbol{\wp}_{v}\right)p\left(\boldsymbol{\theta}|\boldsymbol{\wp}_{v}\right)}d\boldsymbol{\theta}\nonumber \\
 & ={\rm \mathbb{E}}_{q_{\boldsymbol{\eta}}}\left[g\left(\boldsymbol{\eta};\boldsymbol{\theta},\boldsymbol{\wp}_{v}\right)\right],\label{eq:approx_stochastic}
\end{align}
where ${\rm \mathbb{E}}_{q_{\boldsymbol{\eta}}}$ represents the expectation
operator over the variational distribution $q_{\boldsymbol{\eta}}\left(\boldsymbol{\theta}|\boldsymbol{\wp}_{v}\right)$,
and $g\left(\boldsymbol{\eta};\boldsymbol{\theta},\boldsymbol{\wp}_{v}\right){\rm =}\textrm{ln}q_{\boldsymbol{\eta}}\left(\boldsymbol{\theta}|\boldsymbol{\wp}_{v}\right)-\left[\textrm{ln}p\left(\boldsymbol{y}|\boldsymbol{\theta},\boldsymbol{\wp}_{v}\right)+\textrm{ln}p\left(\boldsymbol{\theta}|\boldsymbol{\wp}_{v}\right)\right]$.

To update the parameters $\boldsymbol{\eta}$ and $\boldsymbol{\zeta}$,
we can compute the gradient of the objective function $\boldsymbol{L}\left(\boldsymbol{\eta},\boldsymbol{\zeta}\right)$
with respect to the parameters:
\begin{align}
\nabla_{\boldsymbol{\eta}}\boldsymbol{L}\left(\boldsymbol{\eta},\boldsymbol{\zeta}\right) & =vec\left[\zeta_{v}\nabla_{\boldsymbol{\eta}}{\rm \mathbb{E}}_{q_{\boldsymbol{\eta}}}\left[g\left(\boldsymbol{\eta};\boldsymbol{\theta},\boldsymbol{\wp}_{v}\right)\right]\right]_{v=1}^{N_{\zeta}},\label{eq:Grad_yita}\\
\nabla_{\boldsymbol{\zeta}}\boldsymbol{L}\left(\boldsymbol{\eta},\boldsymbol{\zeta}\right) & =vec\left[\boldsymbol{L}\left(\boldsymbol{\eta}|\boldsymbol{\zeta}\right)+\textrm{ln}\zeta_{v}+1-\textrm{ln}\zeta_{v}^{0}\right]_{v=1}^{N_{\zeta}}.\label{eq:Grad_aita}
\end{align}
For the gradient of an expectation in $\nabla_{\boldsymbol{\eta}}\boldsymbol{L}\left(\boldsymbol{\eta},\boldsymbol{\zeta}\right)$,
following the methodology of SSCA \cite{SSCA,PSSCA}, we can obtain
an asymptotically unbiased estimate through random sampling. Additionally,
the iterative smoothing technique employed in SSCA \cite{SSCA,PSSCA}
ensures that, under certain technical conditions, the parameters $\boldsymbol{\eta}$
being optimized converge to a local optimum. Moreover, from the gradient
$\nabla_{\boldsymbol{\zeta}}\boldsymbol{L}\left(\boldsymbol{\eta},\boldsymbol{\zeta}\right)$,
it can be observed that the updates of model weights primarily depend
on $\boldsymbol{L}\left(\boldsymbol{\eta}|\boldsymbol{\zeta}\right)$,
i.e., the performance of variational distribution fitting the posterior
under different models.

The next subsection details the multi-model random sampling strategy.

\subsection{Auto-focusing Sampling}

According to the update process in SSCA, we can randomly sample the
current variational distribution
\begin{equation}
\boldsymbol{\wp}^{\left(b\right)}\sim q_{\boldsymbol{\zeta}}\left(\boldsymbol{\wp}\right),\boldsymbol{\theta}^{\left(b\right)}\sim q_{\boldsymbol{\eta}}\left(\boldsymbol{\theta}|\boldsymbol{\wp}^{\left(b\right)}\right),\label{eq:Sampling}
\end{equation}
and utilize these realizations $\left(\boldsymbol{\wp}^{\left(b\right)},\boldsymbol{\theta}^{\left(b\right)}\right),\forall b=1,...,B$,
to construct an effective approximation of the objective function
gradient in \eqref{eq:Grad_yita}:
\begin{equation}
\nabla_{\boldsymbol{\eta}}{\rm \mathbb{E}}_{q_{\boldsymbol{\eta}}}\left[g\left(\boldsymbol{\eta};\boldsymbol{\theta},\boldsymbol{\wp}_{v}\right)\right]\approx\frac{1}{B}\sum_{b=1}^{B}\nabla_{\boldsymbol{\eta}}g_{\boldsymbol{\eta}}^{\left(b\right)},\label{eq:sample_approx}
\end{equation}
where $\boldsymbol{\wp}^{\left(b\right)}=\boldsymbol{\wp}_{v}$,
\begin{equation}
g_{\boldsymbol{\eta}}^{\left(b\right)}\triangleq\textrm{ln}\frac{q_{\boldsymbol{\eta}}\left(\boldsymbol{\theta}^{\left(b\right)}|\boldsymbol{\wp}_{v}\right)}{p\left(\boldsymbol{y}|\boldsymbol{\theta}^{\left(b\right)},\boldsymbol{\wp}_{v}\right)p\left(\boldsymbol{\theta}^{\left(b\right)}|\boldsymbol{\wp}_{v}\right)}.\label{eq:grad_g_yita}
\end{equation}
If $N_{\zeta}$ parallel models each sample $B$ times as in the single-model
scenario, the total number of samples would be $N_{\zeta}\times B$,
resulting in computational complexity roughly $N_{\zeta}$ times that
of the single-model SPVBI algorithm. However, since each model has
a posterior probability $\zeta_{v}$, with higher probabilities indicating
a closer fit to the true model, we only need to obtain parameters
for the model with the highest probability. By dynamically allocating
samples based on these probabilities, we can prevent excessive computational
costs during iterations.

\begin{figure}[t]
\begin{centering}
\textsf{\includegraphics[scale=0.55]{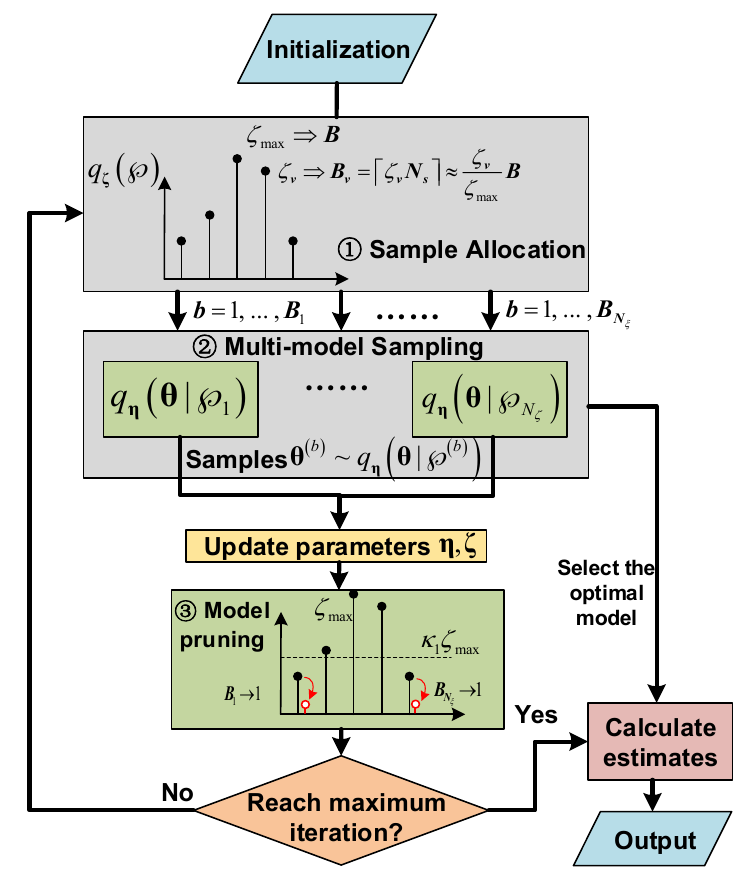}}
\par\end{centering}
\caption{\textsf{\label{Auto-sampling}}Graphical illustration of the auto-focusing
sampling scheme.}
\end{figure}
To achieve this, we propose an auto-focusing sampling scheme, as illustrated
in Fig. \ref{Auto-sampling}. The key idea is to fix the total number
of samples and dynamically allocate them based on model probabilities,
allowing each model to update its parameters accordingly. This approach
concentrates computational resources in the high-probability parameter
space. As iterations progress, the model with the highest probability
gradually dominates, attracting more computational resources, while
others receive fewer samples. Ultimately, the winning model provides
the final parameter estimation.

Specifically, auto-focusing sampling scheme has three key components:
sample allocation, multi-model sampling, and model pruning. (1) First,
the total number of samples per iteration is set to $N_{s}=\left\lceil \frac{B}{\zeta_{max}}\right\rceil $,
where $\zeta_{max}$ is the highest model probability, and $\left\lceil \cdot\right\rceil $
is the ceiling function. Therefore, the number of samples allocated
to each model is $B_{v}=\left\lceil \zeta_{v}N_{s}\right\rceil \approx\frac{\zeta_{v}}{\zeta_{max}}B$.
This ensures that the model with the highest probability consistently
receives at least $B$ samples, similar to single-model SPVBI. Moreover,
during initial model competition, each model starts with sufficient
samples. (2) Subsequently, according to its allocated sample size
$B_{v}$ and variational distribution $q_{\boldsymbol{\eta}}\left(\boldsymbol{\theta}|\boldsymbol{\wp}_{v}\right)$,
each model independently performs random sampling for parameter updates.
(3) Finally, decision mechanisms prune models and establish dominance:
if a model\textquoteright s probability $\zeta_{v}$ drops below $\kappa_{1}\zeta_{max}$,
its sample size $B_{v}$ is reduced to $1$. If $\kappa_{2}\zeta_{max}$
exceeds the second-highest probability $\zeta_{sec,max}$, the top
model's sample size is set to $B$, and all others to $1$. This strategy
quickly prunes less probable models and reduces sample count when
a dominant model emerges. A good choice of $\kappa_{1}$ and $\kappa_{2}$
will be given in the simulation section.

\section{MM-SPVBI Algorithm: Parameter Updates and Convergence}

\subsection{Updating parameter $\mathbf{\boldsymbol{\eta}}$ by SSCA and LS\label{subsec:updatePara}}

In the $t$-th iteration, MM-SPVBI alternates between updating the
parameters $\mathbf{\boldsymbol{\eta}}$ and $\boldsymbol{\zeta}$.
First, by fixing the parameter $\boldsymbol{\zeta}$ as $\boldsymbol{\zeta}^{\left(t\right)}$,
the parameter $\mathbf{\boldsymbol{\eta}}$ in each model can be updated
by solving the following subproblem:
\[
\begin{split}\mathcal{P}_{\boldsymbol{\eta}}:\mathop{\min}\limits_{\boldsymbol{\eta}}\quad & \boldsymbol{L}\left(\boldsymbol{\eta},\boldsymbol{\zeta}^{\left(t\right)}\right)=\sum\limits_{v=1}^{N_{\zeta}}\zeta_{v}^{\left(t\right)}\left[\boldsymbol{L}\left(\boldsymbol{\eta}|\boldsymbol{\zeta}_{v}^{\left(t\right)}\right)+\textrm{ln}\frac{\zeta_{v}^{\left(t\right)}}{\zeta_{v}^{0}}\right]\\
\text{s.t.}\quad & \sum\limits_{n=1}^{N_{p}}w_{k,n}=1,w_{k,n}\in\left[\epsilon,1\right]\quad\forall k,n,\\
 & p_{k,n}\in\left[\hat{\tau}_{k}-\Delta\hat{\tau}_{k}/2,\hat{\tau}_{k}+\Delta\hat{\tau}_{k}/2\right]\quad\forall k,n,\\
 & \mu_{\phi,m}\in\left[0,2\pi\right],\sigma_{\phi,m}^{2}\geq0,\sigma_{\delta,m}^{2}\geq0\quad\forall m.
\end{split}
\]
Considering the signal structure in \ref{eq:matrix_form}, we can
split the parameter $\boldsymbol{\eta}$ into two parts and update
them alternately, namely $\hat{\boldsymbol{\alpha}}^{'}$ and the
other parameters excluding $\hat{\boldsymbol{\alpha}}^{'}$, denoted
as $\boldsymbol{\eta}_{\sim\hat{\boldsymbol{\alpha}}^{'}}\triangleq\left[\mathbf{p}_{1;K},\mathbf{w}_{1;K},\mu_{\delta,1:M},\sigma_{\delta,1:M}^{2},\mu_{\phi,2:M},\sigma_{\phi,2:M}^{2}\right]^{T}$.
For the parameter $\boldsymbol{\eta}_{\sim\hat{\boldsymbol{\alpha}}^{'}}$
of the nonlinear variable $\boldsymbol{\varLambda}$, we can update
it using the SSCA method, similar to the SPVBI algorithm \cite{SPVBI_JIOT}.
For simplicity, $\boldsymbol{\eta}$ below refers to those parameters
within $\boldsymbol{\eta}_{\sim\hat{\boldsymbol{\alpha}}^{'}}$.

\begin{figure}[t]
\begin{centering}
\textsf{\includegraphics[scale=0.7]{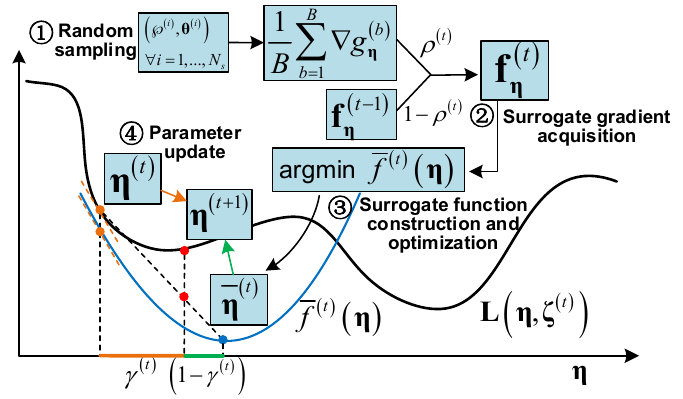}}
\par\end{centering}
\caption{\textsf{\label{SSCA}}Graphical illustration of the stochastic successive
convex approximation.}
\end{figure}

Based on previously obtained samples, we can compute an estimate of
the gradient of the non-convex objective function for each model.
\begin{equation}
\nabla_{\boldsymbol{\eta}}\boldsymbol{L}\left(\boldsymbol{\eta},\boldsymbol{\zeta}^{\left(t\right)}\right)=\zeta_{v}^{\left(t\right)}\nabla_{\boldsymbol{\eta}}\boldsymbol{L}\left(\boldsymbol{\eta}|\boldsymbol{\zeta}_{v}^{\left(t\right)}\right)\approx\frac{\zeta_{v}^{\left(t\right)}}{B_{v}}\sum_{b=1}^{B_{v}}\nabla_{\boldsymbol{\eta}}g_{\boldsymbol{\eta}}^{\left(t,b\right)},\label{eq:grad_yita}
\end{equation}
where $g_{\boldsymbol{\eta}}^{\left(t,b\right)}$ is the sample objective
function as defined in \eqref{eq:grad_g_yita}, $\nabla_{\boldsymbol{\eta}}g_{\boldsymbol{\eta}}^{\left(t,b\right)}$
is its the gradient and the specific expression of the gradient for
the parameters $\mathbf{p}_{k}$, $\mathbf{w}_{k}$, $\mu_{\delta,m}/\mu_{\phi,m}$,
and $\sigma_{\delta,m}^{2}/\sigma_{\phi,m}^{2}$ can be referenced
in Appendix \ref{subsec:grad_1}-\ref{subsec:grad_4}. Next, we can
smooth the gradient $\nabla_{\boldsymbol{\eta}}\boldsymbol{L}\left(\boldsymbol{\eta},\boldsymbol{\zeta}^{\left(t\right)}\right)$
with the historical gradient $\mathbf{\boldsymbol{{\rm f}}}_{\boldsymbol{\eta}}^{\left(t-1\right)}$
under the weighting of a decreasing step size $\rho^{\left(t\right)}$
and we set $\mathbf{\boldsymbol{{\rm f}}}_{\boldsymbol{\eta}}^{\left(-1\right)}=\mathbf{0}$.
\begin{align}
\mathbf{\boldsymbol{{\rm f}}}_{\boldsymbol{\eta}}^{\left(t\right)} & =\left(1-\rho^{\left(t\right)}\right)\mathbf{\boldsymbol{{\rm f}}}_{\boldsymbol{\eta}}^{\left(t-1\right)}+\rho^{\left(t\right)}\nabla_{\boldsymbol{\eta}}\boldsymbol{L}\left(\boldsymbol{\eta},\boldsymbol{\zeta}^{\left(t\right)}\right).\label{eq:f}
\end{align}
Then, guided by the smoothed gradient $\mathbf{\boldsymbol{{\rm f}}}_{\boldsymbol{\eta}}^{\left(t\right)}$,
we construct a series of convex quadratic surrogate functions whose
derivatives can match those of the objective function in the current
iteration, as illustrated by the blue parabola in Fig. \ref{SSCA}.
\begin{align}
\overline{f}^{\left(t\right)}\left(\boldsymbol{\eta}\right) & =\left(\mathbf{\boldsymbol{{\rm f}}}_{\boldsymbol{\eta}}^{\left(t\right)}\right)^{T}\left(\boldsymbol{\eta}-\mathbf{\boldsymbol{\eta}}^{\left(t\right)}\right)+\frac{1}{2\varGamma_{\boldsymbol{\eta}}^{\left(t\right)}}\left\Vert \boldsymbol{\eta}-\mathbf{\boldsymbol{\eta}}^{\left(t\right)}\right\Vert ^{2},\label{eq:surrogate_f}
\end{align}
where $\varGamma_{\boldsymbol{\eta}}^{\left(t\right)}$ is an appropriate
positive number. Solving these surrogate functions can yield an intermediate
value,
\begin{equation}
\overline{\boldsymbol{\eta}}^{\left(t\right)}=\mathrm{arg}\mathop{\mathrm{min}}\limits_{\boldsymbol{\eta}}\overline{f}^{\left(t\right)}\left(\boldsymbol{\eta}\right).\label{eq:surr_subproblem}
\end{equation}
As can be seen, the surrogate optimization problems are quadratic
programming, with the same linear/simplex constraints as in problem
$\mathcal{P}_{\boldsymbol{\eta}}$, which are easy to solve. Intermediate
values are then smoothed with historical values through another step
size $\gamma^{\left(t\right)}$\footnote{To ensure the convergence of the algorithm, the step sizes $\rho^{\left(t\right)}$
and $\gamma^{\left(t\right)}$ must satisfy the following conditions
\cite{SPVBI_JIOT,SSCA}: $\rho^{\left(t\right)}\rightarrow0$, $\sum_{t}\rho^{\left(t\right)}=\infty$,
$\sum_{t}\left(\rho^{\left(t\right)}\right)^{2}<\infty$, $\underset{t\rightarrow\infty}{\lim}\gamma^{\left(t\right)}/\rho^{\left(t\right)}=0$.
A typical choice of $\rho^{\left(t\right)},\gamma^{\left(t\right)}$
is $\rho^{\left(t\right)}=\mathcal{O}\left(t^{-\kappa_{1}}\right)$,
$\gamma^{\left(t\right)}=\mathcal{O}\left(t^{-\kappa_{2}}\right)$,
where $0.5<\kappa_{1}<\kappa_{2}\leq1$.}:
\begin{align}
\boldsymbol{\eta}^{\left(t+1\right)} & =\left(1-\gamma^{\left(t\right)}\right)\mathbf{\boldsymbol{\eta}}^{\left(t\right)}+\gamma^{\left(t\right)}\overline{\boldsymbol{\eta}}^{\left(t\right)}.\label{eq:yita_update}
\end{align}
Based on this process, we continuously update the parameter $\mathbf{\boldsymbol{\eta}}^{\left(t\right)}$
by SSCA as shown in Fig. \ref{SSCA}. It can be demonstrated that,
under certain technical conditions, this iteration guarantees convergence
to a local optimal solution of the original non-convex stochastic
optimization problem $\mathcal{P}_{2}$ \cite{SPVBI_JIOT,SSCA}.

For $\hat{\boldsymbol{\alpha}}^{'}$, when the parameters $\boldsymbol{\eta}_{\sim\hat{\boldsymbol{\alpha}}^{'}}$
are fixed, we find that the maximum likelihood solution $\boldsymbol{\alpha}_{ML}^{'}$
minimizes the KL divergence:
\begin{align}
 & \underset{\textrm{\ensuremath{\hat{\boldsymbol{\alpha}}^{'}}}}{\textrm{argmin}}\boldsymbol{L}\left(\boldsymbol{\eta}_{\sim\hat{\boldsymbol{\alpha}}^{'}}^{\left(t+1\right)},\textrm{\ensuremath{\hat{\boldsymbol{\alpha}}^{'}}}|\boldsymbol{\zeta}_{v}^{\left(t\right)}\right)\nonumber \\
 & =\underset{\textrm{\ensuremath{\hat{\boldsymbol{\alpha}}^{'}}}}{\textrm{argmin}}\int q_{\boldsymbol{\eta}}\left(\boldsymbol{\theta}|\boldsymbol{\eta}_{\sim\hat{\boldsymbol{\alpha}}^{'}}^{\left(t+1\right)},\textrm{\ensuremath{\hat{\boldsymbol{\alpha}}^{'}}}\right)\textrm{ln}\frac{q_{\boldsymbol{\eta}}\left(\boldsymbol{\theta}|\boldsymbol{\eta}_{\sim\hat{\boldsymbol{\alpha}}^{'}}^{\left(t+1\right)},\textrm{\ensuremath{\hat{\boldsymbol{\alpha}}^{'}}}\right)}{p\left(\boldsymbol{y}|\boldsymbol{\theta}\right)p\left(\boldsymbol{\theta}\right)}d\boldsymbol{\theta}\nonumber \\
 & =\underset{\textrm{\ensuremath{\hat{\boldsymbol{\alpha}}^{'}}}}{\textrm{argmax}}\int\int q\left(\boldsymbol{\varLambda}|\boldsymbol{\eta}_{\sim\hat{\boldsymbol{\alpha}}^{'}}^{\left(t+1\right)}\right)q\left(\textrm{\ensuremath{\boldsymbol{\alpha}^{'}}}|\textrm{\ensuremath{\hat{\boldsymbol{\alpha}}^{'}}}\right)\nonumber \\
 & \cdot\textrm{ln}\frac{p\left(\boldsymbol{y}|\boldsymbol{\varLambda},\textrm{\ensuremath{\boldsymbol{\alpha}^{'}}}\right)p\left(\boldsymbol{\varLambda}\right)}{q_{\boldsymbol{\eta}}\left(\boldsymbol{\theta}|\boldsymbol{\eta}_{\sim\hat{\boldsymbol{\alpha}}^{'}}^{\left(t+1\right)},\textrm{\ensuremath{\hat{\boldsymbol{\alpha}}^{'}}}\right)}d\textrm{\ensuremath{\boldsymbol{\alpha}^{'}}}d\boldsymbol{\varLambda}\nonumber \\
 & =\underset{\textrm{\ensuremath{\hat{\boldsymbol{\alpha}}^{'}}}}{\textrm{argmax}}\int q_{\boldsymbol{\eta}}\left(\boldsymbol{\varLambda}|\boldsymbol{\eta}_{\sim\hat{\boldsymbol{\alpha}}^{'}}^{\left(t+1\right)}\right)\textrm{ln}\frac{p\left(\boldsymbol{y}|\boldsymbol{\varLambda},\textrm{\ensuremath{\hat{\boldsymbol{\alpha}}^{'}}}\right)p\left(\boldsymbol{\varLambda}\right)}{q_{\boldsymbol{\eta}}\left(\boldsymbol{\varLambda}|\boldsymbol{\eta}_{\sim\hat{\boldsymbol{\alpha}}^{'}}^{\left(t+1\right)}\right)}d\boldsymbol{\varLambda}\nonumber \\
 & =\underset{\textrm{\ensuremath{\hat{\boldsymbol{\alpha}}^{'}}}}{\textrm{argmax}}\:\textrm{ln}p\left(\boldsymbol{y}|\boldsymbol{\varLambda},\textrm{\ensuremath{\hat{\boldsymbol{\alpha}}^{'}}}\right)=\boldsymbol{\alpha}_{ML}^{'}.\label{eq:alpha_ML}
\end{align}

Furthermore, due to the linear structure in \ref{eq:matrix_form},
the maximum likelihood solution can be obtained using the LS method:
\begin{equation}
\boldsymbol{\alpha}_{ML}^{'}=\left[\mathbf{D}\left(\hat{\boldsymbol{\varLambda}}\right)^{H}\mathbf{D}\left(\hat{\boldsymbol{\varLambda}}\right)\right]^{-1}\mathbf{D}\left(\hat{\boldsymbol{\varLambda}}\right)^{H}\mathbf{y},\label{eq:LS_for_alpha}
\end{equation}
where $\hat{\boldsymbol{\varLambda}}=\left[\hat{\tau}_{1:K},\hat{\phi}_{2:M}^{'},\hat{\delta}_{1:M}\right]^{T}$.
$\hat{\delta}_{m}=\mu_{\delta,m}^{\left(t+1\right)}$, $\hat{\phi}_{m}^{'}=\mu_{\phi,m}^{\left(t+1\right)}$,
$\hat{\tau}_{k}{\rm =}\mathrm{arg}\mathop{\mathrm{max}}\limits_{p_{k,n}}q_{\boldsymbol{\eta}}^{\left(t+1\right)}\left(\tau_{k};\mathbf{p}_{k},\mathbf{w}_{k}\right)$.
To ensure convergence of the algorithm, we further smooth the solution
by applying
\begin{align}
\hat{\boldsymbol{\alpha}}^{'\left(t+1\right)} & =\left(1-\gamma^{\left(t\right)}\right)\hat{\boldsymbol{\alpha}}^{'\left(t\right)}+\gamma^{\left(t\right)}\boldsymbol{\alpha}_{ML}^{'}.\label{eq:alpha_update}
\end{align}
It is worth noting that this improvement does not affect the convergence
of the MM-SPVBI algorithm, as detailed in Section \ref{subsec:Convergence}.

\subsection{Updating parameter $\boldsymbol{\zeta}$\label{subsec:Update_w}}

After updating the parameter $\mathbf{\boldsymbol{\eta}}$ for each
model, the parameter $\boldsymbol{\zeta}$ can be updated by solving
the following subproblem:
\[
\begin{split}\mathcal{P}_{\boldsymbol{\zeta}}:\mathop{\min}\limits_{\boldsymbol{\zeta}}\quad & \boldsymbol{L}\left(\boldsymbol{\eta}^{\left(t+1\right)},\boldsymbol{\zeta}\right)=\sum\limits_{v=1}^{N_{\zeta}}\zeta_{v}\left[\boldsymbol{L}\left(\boldsymbol{\eta}^{\left(t+1\right)}|\zeta_{v}\right)+\textrm{ln}\frac{\zeta_{v}}{\zeta_{v}^{0}}\right]\\
\text{s.t.}\quad & \sum\limits_{v=1}^{N_{\zeta}}\zeta_{v}=1,\quad\zeta_{v}\in\left[\epsilon,1\right],\quad\forall v.
\end{split}
\]
Here, $\boldsymbol{L}\left(\boldsymbol{\eta}^{\left(t+1\right)}|\zeta_{v}\right)$
can be seen as an expectation operation ${\rm \mathbb{E}}_{q_{\boldsymbol{\eta}}^{\left(t+1\right)}}\left[g\left(\boldsymbol{\eta}^{\left(t+1\right)};\boldsymbol{\theta},\boldsymbol{\wp}_{v}\right)\right]$
according to \eqref{eq:approx_stochastic}, making this a convex stochastic
optimization problem with a simplex constraint regarding the variable
$\boldsymbol{\zeta}$.

Similar to problem $\mathcal{P}_{\boldsymbol{\eta}}$, the variable
$\boldsymbol{\zeta}$ can also be updated using the SSCA method. First,
a simple quadratic surrogate objective function
\begin{align}
\overline{f}^{\left(t\right)}\left(\boldsymbol{\zeta}\right) & =\left(\mathbf{\boldsymbol{{\rm f}}}_{\boldsymbol{\zeta}}^{\left(t\right)}\right)^{T}\left(\boldsymbol{\zeta}-\boldsymbol{\zeta}^{\left(t\right)}\right)+\frac{1}{2\varGamma_{\boldsymbol{\zeta}}^{\left(t\right)}}\left\Vert \boldsymbol{\zeta}-\boldsymbol{\zeta}^{\left(t\right)}\right\Vert ^{2}\label{eq:surrogate_f2}
\end{align}
is constructed, and an intermediate variable $\overline{\boldsymbol{\zeta}}^{\left(t\right)}$
is obtained by minimizing this surrogate subproblem as
\begin{align}
\overline{\boldsymbol{\zeta}}^{\left(t\right)} & =\mathrm{arg}\mathop{\mathrm{min}}\limits_{\boldsymbol{\zeta}}\overline{f}^{\left(t\right)}\left(\boldsymbol{\zeta}\right)\nonumber \\
s.t. & \quad\sum\limits_{v=1}^{N_{\zeta}}\zeta_{v}=1,\quad\zeta_{v}\in\left[\epsilon,1\right],\quad\forall v,\label{eq:surr_subproblem2}
\end{align}
where $\varGamma_{\boldsymbol{\zeta}}^{\left(t\right)}$ is an appropriate
positive number. $\mathbf{\boldsymbol{{\rm f}}}_{\boldsymbol{\zeta}}^{\left(t\right)}$
serves as an approximation of the gradient $\nabla_{\boldsymbol{\zeta}}\boldsymbol{L}\left(\boldsymbol{\eta}^{\left(t+1\right)},\boldsymbol{\zeta}\right)$,
which can be iteratively updated as follows:
\begin{align}
\mathbf{\boldsymbol{{\rm f}}}_{\boldsymbol{\zeta}}^{\left(t\right)} & =\left(1-\rho^{\left(t\right)}\right)\mathbf{\boldsymbol{{\rm f}}}_{\boldsymbol{\zeta}}^{\left(t-1\right)}+\rho^{\left(t\right)}\nabla_{\boldsymbol{\zeta}}\boldsymbol{L}\left(\boldsymbol{\eta}^{\left(t+1\right)},\boldsymbol{\zeta}\right),\label{eq:f2}
\end{align}
where
\begin{align}
 & \nabla_{\boldsymbol{\zeta}}\boldsymbol{L}\left(\boldsymbol{\eta}^{\left(t+1\right)},\boldsymbol{\zeta}\right)=vec\left[\boldsymbol{L}\left(\boldsymbol{\eta}^{\left(t+1\right)}|\boldsymbol{\zeta}_{v}\right)+\textrm{ln}\frac{\zeta_{v}}{\zeta_{v}^{0}}+1\right]_{v=1}^{N_{\zeta}}\nonumber \\
 & \approx vec\left[\frac{1}{B_{\zeta}}\sum_{b=1}^{B_{\zeta}}g_{\boldsymbol{\eta}}^{\left(t+1,b\right)}+\textrm{ln}\zeta_{v}+1-\textrm{ln}\zeta_{v}^{0}\right]_{v=1}^{N_{\zeta}}.\label{eq:grad_aita}
\end{align}
Additionally, it holds that $\mathbf{\boldsymbol{{\rm f}}}_{y_{j}}^{\left(-1\right)}=\boldsymbol{0}$.
Finally, $\boldsymbol{\zeta}$ is updated as follows:
\begin{align}
\boldsymbol{\zeta}^{\left(t+1\right)} & =\left(1-\gamma^{\left(t\right)}\right)\boldsymbol{\zeta}^{\left(t\right)}+\gamma^{\left(t\right)}\overline{\boldsymbol{\zeta}}^{\left(t\right)}.\label{eq:aita_update}
\end{align}

\subsection{Convergence Analysis\label{subsec:Convergence}}

In this subsection, we show that the proposed MM-SPVBI algorithm can
be guaranteed to converge to a stationary solution of $\mathcal{P}_{2}$.

Compared to the original single-model SPVBI algorithm in \cite{SPVBI_JIOT},
there are three main differences: the adoption of a hybrid posterior
approximation \eqref{eq:hybrid_post}, the updating of the variable
$\textrm{\ensuremath{\hat{\boldsymbol{\alpha}}^{'}}}$ using the LS
method, and the auto-focusing sampling across multiple models. The
most significant challenge in the proof lies in ensuring that the
optimal closed-form update for certain variables (i.e., $\textrm{\ensuremath{\hat{\boldsymbol{\alpha}}^{'}}}$)
does not affect the convergence of the alternating updates within
the SSCA framework. Below, we present the main convergence results
directly. The detailed proof can be found in Appendix \ref{subsec:Proof-of-Theorem}.
\begin{thm}
[Convergence of MM-SPVBI]\label{thm:Convergence-of-MMSPVBI}Starting
from a feasible initial point $\boldsymbol{\eta}^{\left(0\right)},\boldsymbol{\zeta}^{\left(0\right)}$,
let $\left\{ \boldsymbol{\eta}^{\left(t\right)},\boldsymbol{\zeta}^{\left(t\right)}\right\} _{t=1}^{\infty}$
denote the iterates generated by Algorithm \ref{alg:MM-SPVBI}. Then
every limiting point $\boldsymbol{\eta}^{*},\boldsymbol{\zeta}^{*}$
of $\left\{ \boldsymbol{\eta}^{\left(t\right)},\boldsymbol{\zeta}^{\left(t\right)}\right\} _{t=1}^{\infty}$
is a stationary point of optimization problem $\mathcal{P}_{2}$ almost
surely.
\end{thm}

\subsection{Summary of the MM-SPVBI Algorithm}

Through continuous alternating updates of $\mathbf{\boldsymbol{\eta}}$
and $\boldsymbol{\zeta}$, the proposed MM-SPVBI algorithm can be
guaranteed to converge to a stationary point of problem $\mathcal{P}_{2}$,
as demonstrated in Section \ref{subsec:Convergence}. Subsequently,
we can take the particle position with the highest probability or
the weighted sum of the particles as the final estimate of the delay
$\tau_{k}$, which are the approximate MAP estimate and the minimum
mean square error (MMSE) estimate, respectively. Additionally, the
means $\mu_{\delta,m}$ and $\mu_{\phi,m}$ can serve as the final
estimates for $\delta_{m}$ and $\phi_{m}^{'}$. The optimal model
can also be selected based on the posterior probabilities $\ensuremath{\boldsymbol{\wp}^{*}=\mathrm{arg}\mathop{\mathrm{max}}\limits_{\boldsymbol{\wp}}\:}q_{\boldsymbol{\zeta}}\left(\boldsymbol{\wp}\right)$.
In summary, the overall algorithmic process of MM-SPVBI is summarized
in Algorithm \ref{alg:MM-SPVBI}.
\begin{algorithm}[tbh]
\caption{\label{alg:MM-SPVBI}Multi-model Stochastic Particle VBI Algorithm}

\textbf{Input}: Measurement data, coarse estimates, prior $p\left(\hat{\boldsymbol{\theta}}|\boldsymbol{\wp}\right)p\left(\boldsymbol{\wp}\right)$,
and $\left\{ \rho^{\left(t\right)},\gamma^{\left(t\right)}\right\} $.

\textbf{Initialization}: Split different delay paths to construct
multi-models $\mathcal{M}$, initialize the hybrid variational distribution
$q_{\boldsymbol{\eta}}\left(\boldsymbol{\theta}|\boldsymbol{\wp}\right)$.

\textbf{While not converge do} $(t\rightarrow$$\infty)$

$\quad$ \textbf{Parallel execution for all $N_{\zeta}$ models:}

$\quad\quad$Auto-focus sampling, see Fig. \ref{Auto-sampling};

$\quad\quad$Updating $\boldsymbol{\eta}_{\sim\hat{\boldsymbol{\alpha}}^{'}}$
once via SSCA based on \eqref{eq:f}-\eqref{eq:yita_update};

$\quad\quad$Updating $\textrm{\ensuremath{\hat{\boldsymbol{\alpha}}^{'}}}$
once via LS based on \eqref{eq:LS_for_alpha}-\eqref{eq:alpha_update};

$\quad\quad$Updating $\boldsymbol{\zeta}$ once via SSCA based on
\eqref{eq:surrogate_f2}-\eqref{eq:aita_update}.

\textbf{end}

\textbf{Output}: Identify the best model $\boldsymbol{\wp}^{*}$ with
the maximum weight $\zeta_{v}^{\left(\textrm{max}\right)}$, and extract
posterior estimates $\mathbf{\boldsymbol{\theta}}^{*}$ from the approximate
posterior distribution in the optimal model.
\end{algorithm}

\section{Numerical Simulation and Performance Analysis}

In this section, simulations are conducted to showcase the performance
of the proposed algorithm. We compare its results against the following
baseline algorithms:

\textbf{1) WR-MUSIC} \cite{WR-MUSIC}: As a subspace-based multiband
algorithm, WR-MUSIC can be adopted in the coarse estimation stage
of SPVBI algorithm to narrow the estimation range.

\textbf{2) SPVBI} \cite{SPVBI_JIOT}: Unlike the single-model approach
in the MM-SPVBI algorithm, the original SPVBI employs a full-particle
posterior approximation and fixed-sample-size sampling.

\textbf{3) RJMCMC} \cite{RJMCMC1999}: RJMCMC extends traditional
MCMC methods by enabling trans-dimensional sampling across models
with different numbers of parameters, making it a powerful tool for
Bayesian model selection and averaging.

\textbf{4) Turbo Bayesian inference (Turbo BI)} \cite{wyb_MB}: The
Turbo-BI algorithm employs a hierarchical sparse model based on support
vectors. By iteratively alternating between parameters and support
vectors within the turbo framework, it simultaneously obtains posterior
estimates of both the model order and target parameters.

\textbf{5) Two-Stage Global Estimation (TSGE)} \cite{wyb_MB}: The
TSGE algorithm adopts a two-stage estimation approach: in the stage
1, it uses the Turbo-BI algorithm to identify the model order, and
in the stage 2, it applies traditional methods such as particle swarm
optimization (PSO) and LS to search the high-dimensional parameter
space.

In the simulation, the received signals originate from two non-adjacent
frequency bands, each with a $20\ \textrm{MHz}$ bandwidth, with initial
frequencies of $2.4\ \textrm{GHz}$ and $2.6\ \textrm{GHz}$, respectively.
The subcarrier spacing is $78.125\ \textrm{KHz}$. To assess the algorithm's
resolution limits, we examine low-SNR scenarios with three closely
spaced paths ($\tau_{1}$, $\tau_{2}$, $\tau_{3}$), where $\Delta\tau_{1,2}=30\sim40\ \textrm{ns}$
(below the $50\ \textrm{ns}$ theoretical resolution of $20\ \textrm{MHz}$
bandwidth). Each path has complex gain $\left|\alpha_{k}\right|=0.5$
with uniformly distributed phase $\left(0,2\pi\right)$, while subband
initial phase $\phi_{m}$ and timing synchronization error $\delta_{m}\sim\mathcal{N}\left(0,0.01\ \text{ns}^{2}\right)$
are randomly initialized. Results are averaged over $500$ trials.

Under the aforementioned settings, due to the overlap between $\tau_{1}$
and $\tau_{2}$, only two candidate delay paths can be roughly identified.
Therefore, three candidate models need to be maintained in parallel
for the MM-SPVBI algorithm. This includes the unsplit model and two
models with each delay path split. In each model, each variable is
equipped with $N_{p}=10$ particles, and the mini-batch size $B_{v}$
is set to $10$. The pruning coefficients for the models are $\kappa_{1}=\kappa_{2}=0.5$.
The step sizes are set as follows: $\rho^{\left(t\right)}=5/\left(24+t\right)^{0.51},\rho^{\left(0\right)}=1$;
$\gamma^{\left(t\right)}=5/\left(19+t\right)^{0.55},\gamma^{\left(0\right)}=1$.

\subsection{Target Parameter Estimation Error}

\begin{figure}[t]
\begin{centering}
\textsf{\includegraphics[scale=0.45]{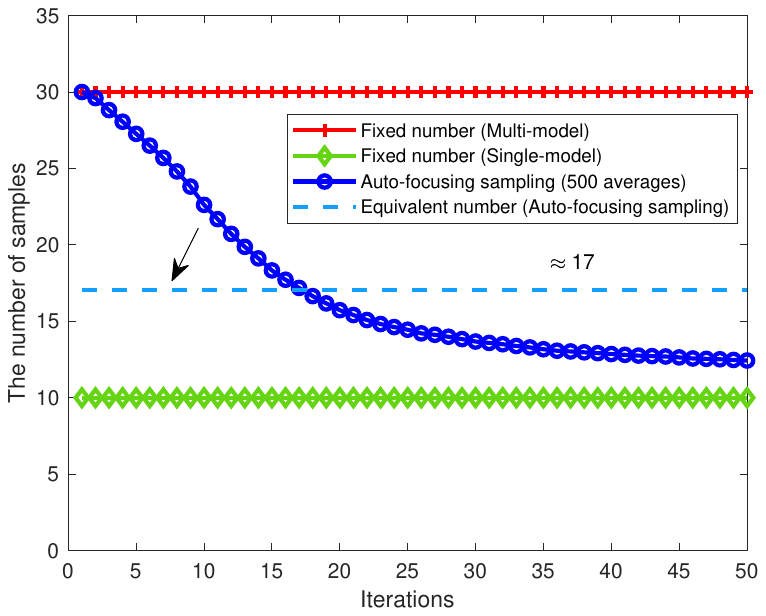}}
\par\end{centering}
\caption{\textsf{\label{Sampling_compare}}Iteration of the number of samples
for different sampling strategies.}
\end{figure}

We first evaluated the iteration behavior of the sample size under
different sampling methods. As depicted in Fig. \ref{Sampling_compare},
if a fixed sample size (e.g., $10$ samples) is assigned uniformly
to variables within a single model, the number of samples for multiple
models will rapidly increase with the degree of parallelism. However,
if we employ an auto-focusing sampling strategy, it allows for the
rapid pruning of poorly performing models after a certain number of
iterations, reallocating computational resources to the most probable
models. Consequently, even with $3$ parallel models, the equivalent
sampling requirement per step is only $17$ samples, preventing significant
complexity escalation.

\begin{figure}[t]
\begin{centering}
\textsf{\includegraphics[scale=0.45]{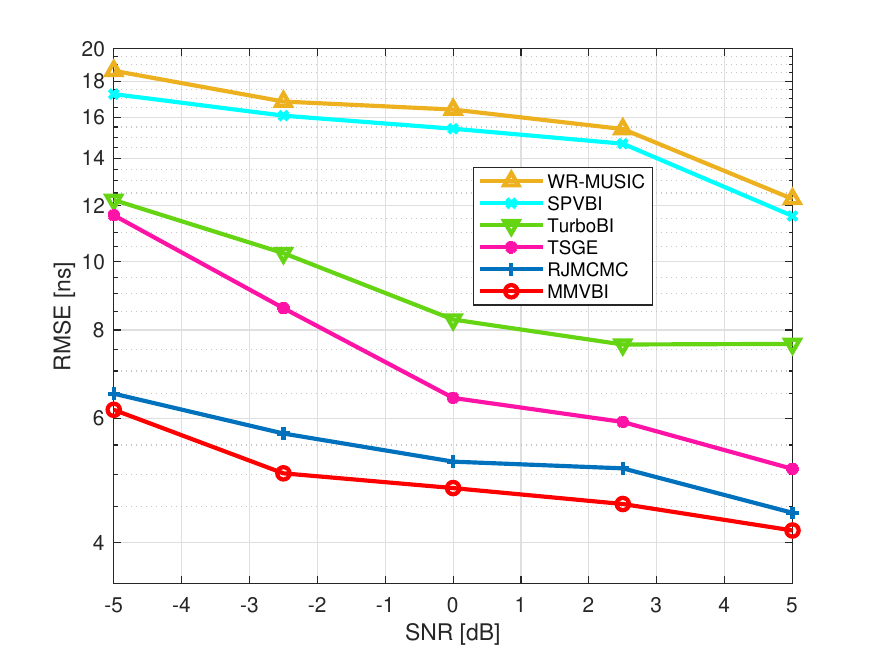}}
\par\end{centering}
\caption{\textsf{\label{RMSE_dSNR}}RMSE of LoS path delay estimation with
respect to the SNR.}
\end{figure}

We then analyze the impact of SNR on LoS path delay estimation accuracy
in the low to mid-SNR range. As shown in Fig. \ref{RMSE_dSNR}, the
performance of most algorithms improves as the SNR increases from
$-5\ \textrm{dB}$ to $5\ \textrm{dB}$. WR-MUSIC and SPVBI perform
suboptimally due to WR-MUSIC's difficulty in distinguishing overlapping
LoS paths under such conditions. Turbo-BI and TSGE offer better resolution
than MUSIC-based algorithms, but as SNR increases, these support vector-based
methods tend to overestimate the number of paths, limiting their accuracy.
TSGE further outperforms Turbo-BI, benefiting from the PSO algorithm
during refined estimation. Notably, the MM-SPVBI algorithm achieves
the lowest root mean square error (RMSE), indicating its superior
accuracy in delay estimation under extreme conditions, driven by more
precise model order selection.

\subsection{Performance of the Detection\label{subsec:Performance}}

\begin{figure}[t]
\begin{centering}
\textsf{\includegraphics[scale=0.45]{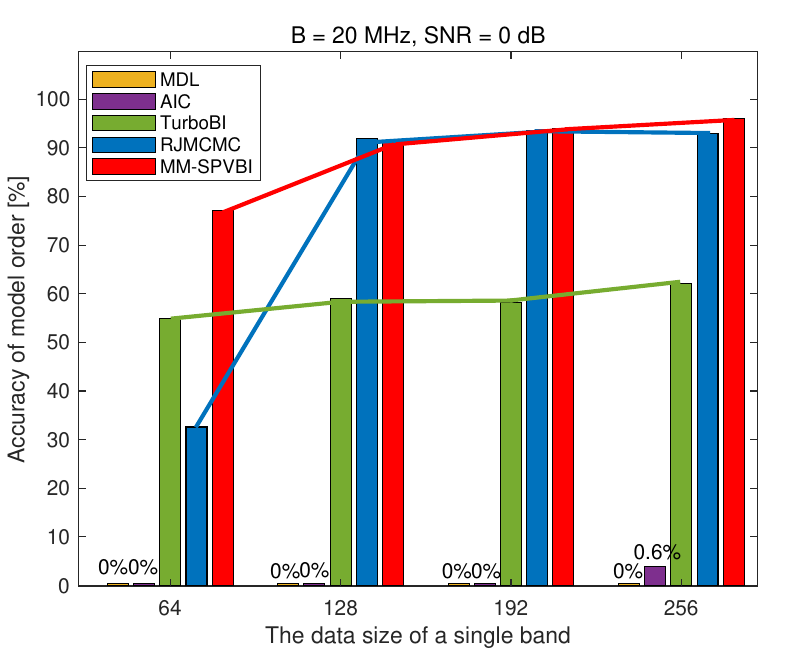}}
\par\end{centering}
\caption{\textsf{\label{Det_datasize}}Detection performance with respect to
the data size.}
\end{figure}

In Fig. \ref{Det_datasize}, we assess the impact of data size (i.e.,
the number of subcarriers adjusted through subcarrier spacing under
fixed bandwidth) on model order accuracy. Traditional criteria nearly
fail under low SNR and closely spaced delays, while the Turbo-BI algorithm's
detection performance hovers around $50\%$. For the RJMCMC and MM-SPVBI,
reducing the data size from $256$ to $128$ causes minimal degradation.
However, at $64$ points, RJMCMC's accuracy drops sharply to $32.6\%$,
while MM-SPVBI maintains $77.2\%$, indicating its strong detection
performance even with shorter data lengths.

\subsection{Algorithm Robustness in Simulated Realistic Channels}

To further evaluate the algorithm\textquoteright s robustness in more
realistic scenarios, we employ the QuaDRiGa platform to generate multiband
CSI samples in an indoor office environment, following the 3GPP NR
standard \cite{3GPP}. The parameters are configured almost identically
to the prior settings. To assess the algorithm\textquoteright s performance
under potential model order misestimation, the SNR is set to $0$
dB.
\begin{figure}[t]
\begin{centering}
\textsf{\includegraphics[scale=0.45]{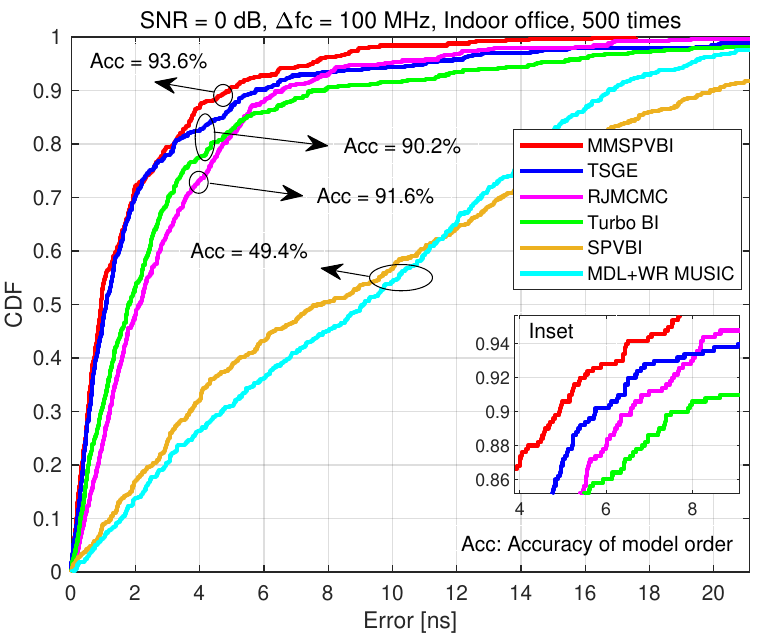}}
\par\end{centering}
\caption{\textsf{\label{CDF}}CDF curves of LoS path delay estimation errors
for different algorithms.}
\end{figure}

In Fig. \ref{CDF}, we plotted the cumulative distribution function
(CDF) curves of the LoS path delay estimation errors for different
algorithms, and annotated the accuracy of each algorithm's estimated
model order. Under such adverse conditions, traditional criteria like
MDL struggle to provide accurate model orders for the SPVBI and WR-MUSIC
algorithms, making it difficult for them to distinguish between closely
spaced delays, $\tau_{1}$ and $\tau_{2}$. This often results in
estimated LoS path delay falling between $\tau_{1}$ and $\tau_{2}$,
leading to significant performance degradation. While SPVBI improves
accuracy during the refined estimation stage, there remains a high
likelihood of bias towards $\tau_{2}$, causing CDF curves exhibit
crossing points. Compared to WR-MUSIC, the Turbo-BI algorithm offers
better model order estimation but tends to overestimate, sometimes
fitting three paths instead of two, reducing precision. The RJMCMC
algorithm demonstrates robust model order determination through trans-dimensional
sampling, but its reliance on histograms from large sample distributions
can lower estimation accuracy. In contrast, the MM-SPVBI algorithm
systematically excludes suboptimal branches and efficiently fits the
posterior marginal distribution, achieving high estimation accuracy
and model precision.

\subsection{Analysis and Comparison of Computational Complexity}

In this subsection, we will focus on analyzing the complexity of the
proposed algorithm and present the complexity results for other algorithms.

For MM-SPVBI, the hybrid posterior approximation and auto-focusing
sampling reduce the parameter count from $2J_{1}N_{p}$ to $J_{3}=2\left(KN_{p}+\left(2M-1\right)\right)$,
with $B_{eq}$ representing the equivalent sample size. The quadratic
programming complexity $\mathcal{O}\left(N_{p}^{3}\right)$ (with
$N_{p}=10$) and LS solution complexity for $\boldsymbol{\alpha}^{'}$
are negligible. Thus, the overall computational complexity reduces
to $\mathcal{O}\left(T_{3}B_{eq}F_{grad}J_{3}\right)$, where $F_{grad}$
denotes the average number of floating-point operations (FLOPs) required
for gradient computation.

For the RJMCMC algorithm \cite{RJMCMC1999}, the overall computational
complexity is $\mathcal{O}\left(T_{1}J_{1}\cdot2F_{L}\right)$, where
$T_{1}$ is the number of iterations, $J_{1}$ is the total number
of variables, and $F_{L}$ denotes the average number of FLOPs required
to compute the likelihood function.

For the WR-MUSIC and SPVBI algorithm, their computational complexities
are as shown in Table \ref{Order_Compare}, where $T_{3}$ is the
number of iterations, and $J_{2}$ is the total number of variables.
A detailed complexity analysis can be found in \cite{SPVBI_JIOT}.

For the Turbo-BI algorithm \cite{wyb_MB}, in the E-step, the complexity
is $\mathcal{O}\left(MN_{m}L^{2}F_{H}\right)$, where $F_{H}$ represents
the average number of FLOPs required to compute a single element of
the observation matrix, and $L$ is the number of offline grid points.
In the M-step, the complexity is approximately $R_{b}F_{L}$, where
$R_{b}$ is the number of iterations for searching the optimal step
size. The total number of turbo iterations is denoted as $T_{4}$.
Therefore, the overall computational complexity is $\mathcal{O}\left(T_{4}\left(MN_{m}L^{2}F_{H}+R_{b}F_{L}\right)\right)$.
For the TSGE algorithm \cite{SPVBI_JIOT,wyb_MB}, the computational
complexity is $\mathcal{O}\left(T_{2}MN_{m}\left(KF_{H}+K^{2}\right)Q_{p}\right)$,
where $Q_{p}$ is the number of particles taken and $T_{2}$ is the
typical number of iterations.

In Table \ref{Order_Compare}, we summarize the complexity order of
different algorithms\textcolor{blue}{{} }in descending order, and numerically
compare them under a typical setting as follows: $T_{1}=2\times10^{4}$,
$J_{1}=6$, $F_{L}=6400$; $T_{2}=500$, $M=2$, $N_{m}=256$, $K=3$,
$F_{H}=10$, $Q_{p}=100$; $T_{3}=35$, $J_{2}=12$, $N_{p}=10$,
$B=10$, $F_{grad}=5700$; $T_{4}=50$, $L=40$, $R_{b}=14$; $B_{eq}=17$,
$J_{3}=66$.
\begin{table}[t]
\caption{\label{Order_Compare}COMPARISON OF THE COMPLEXITY ORDER FOR DIFFERENT
ALGORITHMS.}

\centering{}%
\begin{tabular}{|c|c|c|}
\hline
Algorithms & Complexity order & Typical values\tabularnewline
\hline
RJMCMC & $\mathcal{O}\left(T_{1}J_{1}\cdot2F_{L}\right)$ & $1.54\times10^{9}$\tabularnewline
\hline
TSGE & $\mathcal{O}\left(T_{2}MN_{m}\left(KF_{H}+K^{2}\right)Q_{p}\right)$ & $9.98\times10^{8}$\tabularnewline
\hline
SPVBI & $\mathcal{O}\left(T_{3}2J_{2}\left(N_{p}BF_{grad}+N_{p}^{3}\right)\right)$ & $4.80\times10^{8}$\tabularnewline
\hline
Turbo-BI & $\mathcal{O}\left(T_{4}\left(MN_{m}L^{2}F_{H}+R_{b}F_{L}\right)\right)$ & $4.14\times10^{8}$\tabularnewline
\hline
WR-MUSIC & $\mathcal{O}\left(MN_{m}^{3}+M\left(2N_{m}-1\right)^{3}\right)$ & $3.00\times10^{8}$\tabularnewline
\hline
MM-SPVBI & $\mathcal{O}\left(T_{3}B_{eq}F_{grad}J_{3}\right)$ & $2.24\times10^{8}$\tabularnewline
\hline
\end{tabular}
\end{table}

Compared to the single-model SPVBI with a fixed model order, the MM-SPVBI
algorithm, which updates multiple VBI models in parallel, actually
reduces complexity. This is due to the efficiencies gained from auto-focusing
sampling and hybrid posterior approximation.

\section{Conclusions}

In this paper, we propose a multi-model SPVBI algorithm to address
joint model selection and parameter estimation for multiband delay
estimation under adverse conditions. The MM-SPVBI algorithm extends
single-model SPVBI in two key ways: it updates multiple candidate
models in parallel, improving robustness in identifying significant
delay overlaps, and it enhances the variety of variational distributions
for better posterior approximation. To manage complexity, an auto-focusing
sampling strategy prunes unlikely models, efficiently directing resources
toward the optimal model. Simulations demonstrate the algorithm's
advantages in detection performance, estimation accuracy, and computational
efficiency. Future work will explore random set theory for more complex
joint model selection and parameter estimation, as well as temporal
correlations for Bayesian inference and tracking. Additionally, we
plan to conduct experimental validation using real-world data in more
complex multipath environments to further verify the algorithm's performance.

\appendix

\subsection{Proof of Theorem\ref{thm:Convergence-of-MMSPVBI} \label{subsec:Proof-of-Theorem}}

We first establish that the hybrid posterior approximation and LS-based
update of $\hat{\boldsymbol{\alpha}}^{'}$ preserve convergence. While
the posterior distributions for $\phi_{m}^{'}$ and $\delta_{m}$
transition from particle-based to Gaussian forms, and while we introduce
the new model variable $\boldsymbol{\zeta}$, the SSCA-based solution
of subproblems $\mathcal{P}_{\boldsymbol{\eta}}$ and $\mathcal{P}_{\boldsymbol{\zeta}}$
maintains convergence. Thus, similar to the convergence proof in \cite{SPVBI_JIOT},
after performing steps such as random sampling, approximate gradient
calculation, gradient smoothing, and parameter updating, convergence
to a stationary point can be ensured under standard step size conditions
($\rho^{t},\gamma^{t}$).

Moreover, the LS-based update of $\hat{\boldsymbol{\alpha}}^{'}$
with smoothing in \eqref{eq:alpha_update} represents a special SSCA
instance, preserving convergence guarantees. Unlike the subproblems
for $\boldsymbol{\eta}_{\sim\hat{\boldsymbol{\alpha}}^{'}}$ and $\boldsymbol{\zeta}$,
when $\boldsymbol{\eta}_{\sim\hat{\boldsymbol{\alpha}}^{'}}$ and
$\boldsymbol{\zeta}$ are fixed, the subproblem $\underset{\textrm{\ensuremath{\hat{\boldsymbol{\alpha}}^{'}}}}{\textrm{min}}\boldsymbol{L}\left(\boldsymbol{\eta}_{\sim\hat{\boldsymbol{\alpha}}^{'}}^{\left(t+1\right)},\textrm{\ensuremath{\hat{\boldsymbol{\alpha}}^{'}}}|\boldsymbol{\zeta}_{v}^{\left(t\right)}\right)$
with respect to $\hat{\boldsymbol{\alpha}}^{'}$ is strongly convex.
This is due to the fact that
\begin{equation}
\begin{split} & \nabla_{\textrm{\ensuremath{\hat{\boldsymbol{\alpha}}^{'}}}}^{2}\boldsymbol{L}\left(\boldsymbol{\eta}_{\sim\hat{\boldsymbol{\alpha}}^{'}}^{\left(t+1\right)},\textrm{\ensuremath{\hat{\boldsymbol{\alpha}}^{'}}}|\boldsymbol{\zeta}_{v}^{\left(t\right)}\right)\\
 & =\nabla_{\textrm{\ensuremath{\hat{\boldsymbol{\alpha}}^{'}}}}^{2}\left\{ {\rm \mathbb{E}}_{q_{\boldsymbol{\eta}}\left(\boldsymbol{\varLambda}\right)}\left[-\textrm{ln}p\left(\boldsymbol{y}|\boldsymbol{\varLambda},\textrm{\ensuremath{\hat{\boldsymbol{\alpha}}^{'}}}\right)\right]+C\right\} \\
 & ={\rm \mathbb{E}}_{q_{\boldsymbol{\eta}}\left(\boldsymbol{\varLambda}\right)}\left[\nabla_{\textrm{\ensuremath{\hat{\boldsymbol{\alpha}}^{'}}}}^{2}\textrm{ln}p\left(\boldsymbol{y}|\boldsymbol{\varLambda},\textrm{\ensuremath{\hat{\boldsymbol{\alpha}}^{'}}}\right)\right]\\
 & ={\rm \mathbb{E}}_{q_{\boldsymbol{\eta}}\left(\boldsymbol{\varLambda}\right)}\left[\frac{1}{\sigma_{w}^{2}}\mathbf{D}\left(\boldsymbol{\varLambda}\right)^{H}\mathbf{D}\left(\boldsymbol{\varLambda}\right)\right]\succeq0,
\end{split}
\end{equation}
where $C$ is a constant and the expectation ${\rm \mathbb{E}}_{q_{\boldsymbol{\eta}}\left(\boldsymbol{\varLambda}\right)}\left[\cdot\right]$
is taken with respect to the distribution $q_{\boldsymbol{\eta}}\left(\boldsymbol{\varLambda}|\boldsymbol{\eta}_{\sim\hat{\boldsymbol{\alpha}}^{'}}^{\left(t+1\right)}\right)$.
Moreover, as shown in \eqref{eq:LS_for_alpha}, this strongly convex
function has a closed-form solution $\boldsymbol{\alpha}_{ML}^{'}$.
Therefore, it is unnecessary to construct an additional convex surrogate
function; instead, we can directly use it as the surrogate function
$\overline{f}^{\left(t\right)}\left(\hat{\boldsymbol{\alpha}}^{'}\right)=\boldsymbol{L}\left(\boldsymbol{\eta}_{\sim\hat{\boldsymbol{\alpha}}^{'}}^{\left(t+1\right)},\textrm{\ensuremath{\hat{\boldsymbol{\alpha}}^{'}}}|\boldsymbol{\zeta}_{v}^{\left(t\right)}\right)$
and obtain the optimal value for $\hat{\boldsymbol{\alpha}}^{'}$
as in \eqref{eq:LS_for_alpha}.

The choice of $\boldsymbol{L}\left(\boldsymbol{\eta}_{\sim\hat{\boldsymbol{\alpha}}^{'}}^{\left(t+1\right)},\textrm{\ensuremath{\hat{\boldsymbol{\alpha}}^{'}}}|\boldsymbol{\zeta}_{v}^{\left(t\right)}\right)$
as the surrogate function for $\hat{\boldsymbol{\alpha}}^{'}$ optimization
naturally satisfies the asymptotic consistency condition (Lemma 4,
\cite{SPVBI_JIOT}):
\begin{align}
\left\Vert \nabla_{\hat{\boldsymbol{\alpha}}^{'}}\hat{f}\left(\hat{\boldsymbol{\alpha}}^{'*}\right)-\nabla_{\hat{\boldsymbol{\alpha}}^{'}}\boldsymbol{L}\left(\boldsymbol{\eta}^{*},\boldsymbol{\zeta}^{*}\right)\right\Vert  & =0,
\end{align}
where $\mathop{\lim}\limits_{t\to\infty}\overline{f}^{\left(t\right)}\left(\hat{\boldsymbol{\alpha}}^{'}\right)\triangleq\hat{f}\left(\hat{\boldsymbol{\alpha}}^{'}\right)$
and $\hat{\boldsymbol{\alpha}}^{'*}$ is a limit point. Therefore,
the smoothed LS solution for $\textrm{\ensuremath{\hat{\boldsymbol{\alpha}}^{'}}}$
can also be viewed as the solution provided by the SSCA method. Since
$\boldsymbol{\eta}_{\sim\hat{\boldsymbol{\alpha}}^{'}}$, $\hat{\boldsymbol{\alpha}}^{'}$,
and $\boldsymbol{\zeta}$ are independent of each other, the proposed
updating scheme can be viewed as an alternating optimization (AO)
within the SSCA framework. Similar to the approach used in \cite{SPVBI_JIOT}
to prove the alternating updates in SSCA, each iteration of Algorithm
\ref{alg:MM-SPVBI} corresponds to optimizing the following surrogate
function, $\overline{f}^{\left(t\right)}\left(\boldsymbol{\eta},\boldsymbol{\zeta}\right)=\overline{f}^{\left(t\right)}\left(\boldsymbol{\eta}\right)+\overline{f}^{\left(t\right)}\left(\boldsymbol{\zeta}\right).$
Using the properties of the surrogate function and the similar analysis
as in the proof of (\cite{SPVBI_JIOT}, Theorem 1), it is straightforward
to prove that $\mathop{\lim}\limits_{t\to\infty}\overline{f}^{\left(t\right)}\left(\boldsymbol{\eta}^{*},\boldsymbol{\zeta}^{*}\right)\triangleq\hat{f}\left(\boldsymbol{\eta}^{*},\boldsymbol{\zeta}^{*}\right)$,
and $\nabla\hat{f}\left(\boldsymbol{\eta}^{*},\boldsymbol{\zeta}^{*}\right)=\nabla\boldsymbol{L}\left(\boldsymbol{\eta},\boldsymbol{\zeta}\right)$.
Therefore, $\left(\boldsymbol{\eta}^{*},\boldsymbol{\zeta}^{*}\right)$
also satisfies the KKT conditions of the original problem $\mathcal{P}_{2}$.

We now establish that auto-focusing sampling preserves convergence.
The model selection subproblem $\mathcal{P}_{\boldsymbol{\zeta}}$
updates model weights $\zeta_{v}$ based on performance, guiding the
sampling strategy to dynamically allocate resources: dominant models
receive more samples while all models maintain at least one. This
ensures each parallel SPVBI instance can converge to local optima,
with the final solution selected from these candidates - necessarily
a local optimum of $\mathcal{P}_{2}$ and potentially the global optimum.
By systematically exploring a broader parameter space than SPVBI or
TSGE, while maintaining computational efficiency, our multi-model
approach achieves superior global optimization capability.

\subsection{Gradient of Particle Position\label{subsec:grad_1}}

According to
\begin{equation}
\nabla_{\boldsymbol{\eta}}\boldsymbol{L}\left(\boldsymbol{\eta},\boldsymbol{\zeta}^{\left(t\right)}\right)=\zeta_{v}^{\left(t\right)}\nabla_{\boldsymbol{\eta}}\boldsymbol{L}\left(\boldsymbol{\eta}|\boldsymbol{\zeta}_{v}^{\left(t\right)}\right)\approx\frac{\zeta_{v}^{\left(t\right)}}{B_{v}}\sum_{b=1}^{B_{v}}\nabla g_{\boldsymbol{\eta}}^{\left(t,b\right)},
\end{equation}
we can obtain the gradient $\nabla_{\boldsymbol{\eta}}\boldsymbol{L}\left(\boldsymbol{\eta},\boldsymbol{\zeta}^{\left(t\right)}\right)$
of the objective function with respect to the parameter $\boldsymbol{\eta}$
by computing $\nabla_{\boldsymbol{\eta}}\boldsymbol{L}\left(\boldsymbol{\eta}|\boldsymbol{\zeta}_{v}^{\left(t\right)}\right)$.

For the gradient of particle position with respect to the variable
$\tau_{k}$, we first isolate the variational distribution $q_{\tau_{k}}$
as follows:
\begin{align}
 & \nabla_{\mathbf{p}_{k}}\boldsymbol{L}\left(\boldsymbol{\eta}|\boldsymbol{\zeta}_{v}^{\left(t\right)}\right)\nonumber \\
 & \triangleq\nabla_{\mathbf{p}_{k}}\int q_{\boldsymbol{\eta}}\left(\boldsymbol{\theta}|\boldsymbol{\wp}_{v}^{\left(t\right)}\right)\textrm{ln}\frac{q_{\boldsymbol{\eta}}\left(\boldsymbol{\theta}|\boldsymbol{\wp}_{v}^{\left(t\right)}\right)}{p\left(\boldsymbol{y}|\boldsymbol{\theta},\boldsymbol{\wp}_{v}^{\left(t\right)}\right)p\left(\boldsymbol{\theta}|\boldsymbol{\wp}_{v}^{\left(t\right)}\right)}d\boldsymbol{\theta}\nonumber \\
 & =\nabla_{\mathbf{p}_{k}}\int q_{\sim\tau_{k}}\left[q_{\tau_{k}}\textrm{ln}\frac{q_{\tau_{k}}q_{\sim\tau_{k}}}{p\left(\boldsymbol{y}|\boldsymbol{\theta},\boldsymbol{\wp}_{v}^{\left(t\right)}\right)p\left(\boldsymbol{\theta}|\boldsymbol{\wp}_{v}^{\left(t\right)}\right)}d\tau_{k}\right]d\boldsymbol{\theta}_{\sim\tau_{k}}\nonumber \\
 & =\nabla_{\mathbf{p}_{k}}{\rm \mathbb{E}}_{q_{\sim\tau_{k}}}\left[g_{k}^{\left(t\right)}\left(\mathbf{p}_{k};\mathbf{w}_{k}^{\left(t\right)},\boldsymbol{\theta}_{\sim\tau_{k}},\boldsymbol{\wp}_{v}^{\left(t\right)}\right)\right]\nonumber \\
 & ={\rm \mathbb{E}}_{q_{\sim\tau_{k}}}\left[\nabla_{\mathbf{p}_{k}}g_{k}^{\left(t\right)}\left(\mathbf{p}_{k};\mathbf{w}_{k}^{\left(t\right)},\boldsymbol{\theta}_{\sim\tau_{k}},\boldsymbol{\wp}_{v}^{\left(t\right)}\right)\right],
\end{align}
where $q_{\sim\tau_{k}}\coloneqq q_{\boldsymbol{\eta}}\left(\boldsymbol{\theta}_{\sim\tau_{k}}|\boldsymbol{\wp}_{v}^{\left(t\right)}\right)$,
$q_{\tau_{k}}\coloneqq q_{\boldsymbol{\eta}}\left(\tau_{k}|\boldsymbol{\wp}_{v}^{\left(t\right)}\right)$,
and
\begin{align}
 & g_{k}^{\left(t\right)}\left(\mathbf{p}_{k};\mathbf{w}_{k}^{\left(t\right)},\boldsymbol{\theta}_{\sim\tau_{k}},\boldsymbol{\wp}_{v}^{\left(t\right)}\right)\nonumber \\
 & =\sum\limits_{n=1}^{N_{p}}w_{k,n}^{\left(t\right)}\delta\left(\tau_{k}-p_{k,n}\right)\textrm{ln}\frac{q_{\tau_{k}}q_{\sim\tau_{k}}}{p\left(\boldsymbol{y}|\boldsymbol{\theta},\boldsymbol{\wp}_{v}^{\left(t\right)}\right)p\left(\boldsymbol{\theta}|\boldsymbol{\wp}_{v}^{\left(t\right)}\right)}\nonumber \\
 & =\sum\limits_{n=1}^{N_{p}}w_{k,n}^{\left(t\right)}\textrm{ln}\frac{w_{k,n}^{\left(t\right)}}{p\left(\boldsymbol{y}|\mathbf{\boldsymbol{\theta}}_{\sim\tau_{k}},p_{k,n},\boldsymbol{\wp}_{v}^{\left(t\right)}\right)p\left(\boldsymbol{\theta}_{\sim\tau_{k}},p_{k,n}|\boldsymbol{\wp}_{v}^{\left(t\right)}\right)}.
\end{align}
For the outer expectation with respect to $q_{\sim\tau_{k}}$, we
can approximate it through random sampling. Therefore, the gradient
can be expressed as follows:
\begin{align}
\nabla_{\mathbf{p}_{k}} & \boldsymbol{L}\left(\boldsymbol{\eta}|\boldsymbol{\zeta}_{v}^{\left(t\right)}\right)={\rm \mathbb{E}}_{q_{\sim\tau_{k}}}\left[\nabla_{\mathbf{p}_{k}}g_{k}^{\left(t\right)}\left(\mathbf{p}_{k},\mathbf{w}_{k}^{\left(t\right)};\boldsymbol{\theta}_{\sim\tau_{k}},\boldsymbol{\wp}_{v}^{\left(t\right)}\right)\right]\nonumber \\
\approx & \frac{1}{B_{v}}\sum_{b=1}^{B_{v}}\nabla_{\mathbf{p}_{k}}g_{k}^{\left(t,b\right)}\left(\mathbf{p}_{k};\mathbf{w}_{k}^{\left(t\right)}\mathbf{\boldsymbol{\theta}}_{\sim\tau_{k}}^{\left(b\right)},\boldsymbol{\wp}_{v}^{\left(t\right)}\right)_{\mathbf{p}_{k}=\mathbf{p}_{k}^{\left(t\right)}}\nonumber \\
\triangleq & \frac{1}{B_{v}}\sum_{b=1}^{B_{v}}\nabla g_{\mathbf{p}_{k}}^{\left(t,b\right)},\label{eq:app_grad_p}
\end{align}
where
\begin{align}
 & \nabla g_{\mathbf{p}_{k}}^{\left(t,b\right)}\nonumber \\
 & =vec\left\{ -w_{k,n}^{\left(t\right)}\nabla_{p_{k,n}}\left[\textrm{ln}p\left(\boldsymbol{y}|\mathbf{\boldsymbol{\theta}}_{\sim\tau_{k}}^{\left(b\right)},p_{k,n}^{\left(t\right)},\boldsymbol{\wp}_{v}^{\left(t\right)}\right)\right.\right.\nonumber \\
 & \left.\left.+\textrm{ln}p\left(\mathbf{\boldsymbol{\theta}}_{\sim\tau_{k}}^{\left(b\right)},p_{k,n}^{\left(t\right)}|\boldsymbol{\wp}_{v}^{\left(t\right)}\right)\right]\right\} _{n=1}^{N_{p}}\nonumber \\
 & =vec\left\{ -w_{k,n}^{\left(t\right)}\nabla_{p_{k,n}}\textrm{ln}p\left(\boldsymbol{y}|\mathbf{\boldsymbol{\theta}}_{\sim\tau_{k}}^{\left(b\right)},p_{k,n},\boldsymbol{\wp}_{v}^{\left(t\right)}\right)_{p_{k,n}=p_{k,n}^{\left(t\right)}}\right\} _{n=1}^{N_{p}}.\label{eq:app_g_p}
\end{align}
The final equality holds because we assume the prior $\textrm{ln}p\left(\mathbf{\boldsymbol{\theta}}_{\sim\tau_{k}}^{\left(b\right)},p_{k,n}|\boldsymbol{\wp}_{v}^{\left(t\right)}\right)$
is a uniformly discrete distribution within the estimation interval,
resulting in a gradient of zero with respect to the particle position.
For the gradient of the likelihood function with respect to the particle
position $\nabla_{p_{k,n}}\textrm{ln}p\left(\boldsymbol{y}|\mathbf{\boldsymbol{\theta}}_{\sim\tau_{k}}^{\left(b\right)},p_{k,n},\boldsymbol{\wp}_{v}^{\left(t\right)}\right)$,
we have:
\begin{align}
 & \nabla_{p_{k,n}}\textrm{ln}p\left(\boldsymbol{y}|\mathbf{\boldsymbol{\theta}}_{\sim\tau_{k}}^{\left(b\right)},p_{k,n},\boldsymbol{\wp}_{v}^{\left(t\right)}\right)\nonumber \\
 & =\frac{-1}{\sigma_{w}^{2}}\sum\limits_{m=1}^{M}\sum\limits_{n=0}^{N_{m}-1}\frac{\partial\left|y_{m}^{\left(n\right)}-s_{m}^{\left(n\right)}\right|^{2}}{\partial p_{k,n}}\nonumber \\
 & =\frac{-2}{\sigma_{w}^{2}}\sum\limits_{m=1}^{M}\sum\limits_{n=0}^{N_{m}-1}\left[\textrm{Re}\left(s_{m}^{\left(n\right)}-y_{m}^{\left(n\right)}\right)\nabla_{p_{k,n}}\textrm{Re}\left(s_{m}^{\left(n\right)}\right)\right.\nonumber \\
 & \left.+\textrm{Im}\left(s_{m}^{\left(n\right)}-y_{m}^{\left(n\right)}\right)\nabla_{p_{k,n}}\textrm{Im}\left(s_{m}^{\left(n\right)}\right)\right],
\end{align}
where
\begin{align}
\nabla_{p_{k,n}}\textrm{Re}\left(s_{m}^{\left(n\right)}\right) & =2\pi(f_{c,m}^{'}+nf_{s,m})\left|\hat{\alpha}_{k}^{'}\right|sin\left(\vartheta_{k,m,n}\right),\\
\nabla_{p_{k,n}}\textrm{Im}\left(s_{m}^{\left(n\right)}\right) & =-2\pi(f_{c,m}^{'}+nf_{s,m})\left|\hat{\alpha}_{k}^{'}\right|cos\left(\vartheta_{k,m,n}\right),
\end{align}
and $\vartheta_{k,m,n}=\angle\hat{\alpha}_{k}^{'}+\phi_{m}^{'\left(b\right)}-2\pi(f_{c,m}^{'}+nf_{s,m})\tau_{k}^{\left(b\right)}-2\pi\left(nf_{s,m}\right)\delta_{m}^{\left(b\right)}$.

\subsection{Gradient of Particle Weight\label{subsec:grad_2}}

For the gradient of particle weights, according to \eqref{eq:app_grad_p}
and \eqref{eq:app_g_p}, we similarly have:
\begin{align}
\nabla_{\mathbf{w}_{k}} & \boldsymbol{L}\left(\boldsymbol{\eta}|\boldsymbol{\zeta}_{v}^{\left(t\right)}\right)={\rm \mathbb{E}}_{q_{\sim\tau_{k}}}\left[\nabla_{\mathbf{w}_{k}}g_{k}^{\left(t\right)}\left(\mathbf{p}_{k}^{\left(t\right)},\mathbf{w}_{k};\boldsymbol{\theta}_{\sim\tau_{k}},\boldsymbol{\wp}_{v}^{\left(t\right)}\right)\right]\nonumber \\
\approx & \frac{1}{B_{v}}\sum_{b=1}^{B_{v}}\nabla_{\mathbf{w}_{k}}g_{k}^{\left(t,b\right)}\left(\mathbf{p}_{k}^{\left(t\right)},\mathbf{w}_{k};\mathbf{\boldsymbol{\theta}}_{\sim\tau_{k}}^{\left(b\right)},\boldsymbol{\wp}_{v}^{\left(t\right)}\right)_{\mathbf{w}_{k}=\mathbf{w}_{k}^{\left(t\right)}}\nonumber \\
\triangleq & \frac{1}{B_{v}}\sum_{b=1}^{B_{v}}\nabla g_{\mathbf{w}_{k}}^{\left(t,b\right)},\label{eq:app_grad_w}
\end{align}
where
\begin{align}
\nabla & g_{\mathbf{w}_{k}}^{\left(t,b\right)}=vec\left[\textrm{ln}w_{k,n}^{\left(t\right)}+1-\textrm{ln}p\left(\mathbf{\boldsymbol{\theta}}_{\sim\tau_{k}}^{\left(b\right)},p_{k,n}^{\left(t\right)}|\boldsymbol{\wp}_{v}^{\left(t\right)}\right)\right.\nonumber \\
- & \left.\textrm{ln}p\left(\boldsymbol{r}|\mathbf{\boldsymbol{\theta}}_{\sim\tau_{k}}^{\left(b\right)},p_{k,n}^{\left(t\right)},\boldsymbol{\wp}_{v}^{\left(t\right)}\right)\right]_{n=1}^{N_{p}}.\label{eq:app_g_w}
\end{align}

\subsection{Gradient of the Mean of the Gaussian Distribution\label{subsec:grad_3}}

We first have the following result:
\begin{align}
 & \nabla_{\boldsymbol{\eta}}\boldsymbol{L}\left(\boldsymbol{\eta}|\boldsymbol{\zeta}_{v}^{\left(t\right)}\right)\nonumber \\
 & \triangleq\nabla_{\boldsymbol{\eta}}\int q_{\boldsymbol{\eta}}\left(\boldsymbol{\theta}|\boldsymbol{\wp}_{v}^{\left(t\right)}\right)\textrm{ln}\frac{q_{\boldsymbol{\eta}}\left(\boldsymbol{\theta}|\boldsymbol{\wp}_{v}^{\left(t\right)}\right)}{p\left(\boldsymbol{r}|\boldsymbol{\theta},\boldsymbol{\wp}_{v}^{\left(t\right)}\right)p\left(\boldsymbol{\theta}|\boldsymbol{\wp}_{v}^{\left(t\right)}\right)}d\boldsymbol{\theta}_{K}\nonumber \\
 & ={\rm \mathbb{E}}_{q_{\boldsymbol{\eta}}}\ensuremath{\left\{ \left[\nabla_{\boldsymbol{\eta}}\textrm{ln}q_{\boldsymbol{\eta}}\left(\boldsymbol{\theta}|\boldsymbol{\wp}_{v}^{\left(t\right)}\right)\right]\textrm{ln}\frac{q_{\boldsymbol{\eta}}\left(\boldsymbol{\theta}|\boldsymbol{\wp}_{v}^{\left(t\right)}\right)}{p\left(\boldsymbol{r}|\boldsymbol{\theta},\boldsymbol{\wp}_{v}^{\left(t\right)}\right)p\left(\boldsymbol{\theta}|\boldsymbol{\wp}_{v}^{\left(t\right)}\right)}\right\} .}\label{eq:app_grad_lemma}
\end{align}
For the gradient of mean $\mu_{\delta,m}$
\begin{equation}
\nabla_{\mu_{\delta,m}}\textrm{ln}q_{\boldsymbol{\eta}}\left(\boldsymbol{\theta}|\boldsymbol{\wp}_{v}^{\left(t\right)}\right)=\nabla_{\mu_{\delta,m}}\textrm{ln}\mathcal{N}\left(\delta_{m};\mu_{\delta,m},\sigma_{\delta,m}^{2}\right)=\frac{\delta_{m}-\mu_{\delta,m}}{\sigma_{\delta,m}^{2}}.
\end{equation}
Similarly, it can be approximated through random sampling:
\begin{align}
\nabla_{\mu_{\delta,m}}\boldsymbol{L}\left(\boldsymbol{\eta}|\boldsymbol{\zeta}_{v}^{\left(t\right)}\right) & \approx\frac{1}{B_{v}}\sum_{b=1}^{B_{v}}\left\{ \left[\nabla_{\mu_{\delta,m}}\textrm{ln}q_{\boldsymbol{\eta}}\left(\boldsymbol{\theta}^{\left(b\right)}|\boldsymbol{\wp}_{v}^{\left(t\right)}\right)\right]g_{\boldsymbol{\eta}}^{\left(t,b\right)}\right\} \nonumber \\
 & =\frac{1}{B_{v}}\sum_{b=1}^{B_{v}}\left[\ensuremath{\frac{\delta_{m}^{\left(b\right)}-\mu_{\delta,m}^{\left(t\right)}}{\sigma_{\delta,m}^{\left(t\right)2}}}g_{\boldsymbol{\eta}}^{\left(t,b\right)}\right],
\end{align}
where $g_{\boldsymbol{\eta}}^{\left(t,b\right)}\triangleq\textrm{ln}\frac{q_{\boldsymbol{\eta}}\left(\boldsymbol{\theta}^{\left(b\right)}|\boldsymbol{\wp}_{v}^{\left(t\right)}\right)}{p\left(\boldsymbol{r}|\boldsymbol{\theta}^{\left(b\right)},\boldsymbol{\wp}_{v}^{\left(t\right)}\right)p\left(\boldsymbol{\theta}^{\left(b\right)}|\boldsymbol{\wp}_{v}^{\left(t\right)}\right)}$.
In the above expression, taking $\mu_{\delta,m}$ as an example, the
derivation is similar for $\mu_{\phi,m}$.

\subsection{Gradient of the Variance of the Gaussian Distribution\label{subsec:grad_4}}

Similarly, using \eqref{eq:app_grad_lemma}, for the gradient of variance
$\sigma_{\delta,m}^{2}$, we have
\begin{align}
 & \nabla_{\sigma_{\delta,m}^{2}}\textrm{ln}q_{\boldsymbol{\eta}}\left(\boldsymbol{\theta}|\boldsymbol{\wp}_{v}^{\left(t\right)}\right)\nonumber \\
 & =\nabla_{\sigma_{\delta,m}^{2}}\textrm{ln}\mathcal{N}\left(\delta_{m};\mu_{\delta,m},\sigma_{\delta,m}^{2}\right)=\frac{-1}{2\sigma_{\delta,m}^{2}}+\frac{\left(\delta_{m}-\mu_{\delta,m}\right)^{2}}{2\sigma_{\delta,m}^{4}}.
\end{align}
Therefore,
\begin{align}
 & \nabla_{\sigma_{\delta,m}^{2}}\boldsymbol{L}\left(\boldsymbol{\eta}|\boldsymbol{\zeta}_{v}^{\left(t\right)}\right)\nonumber \\
 & \approx\frac{1}{B_{v}}\sum_{b=1}^{B_{v}}\left\{ \left[\nabla_{\sigma_{\delta,m}^{2}}\textrm{ln}q_{\boldsymbol{\eta}}\left(\boldsymbol{\theta}|\boldsymbol{\wp}_{v}^{\left(t\right)}\right)\right]\cdot g_{\boldsymbol{\eta}}^{\left(t,b\right)}\right\} \nonumber \\
 & =\frac{1}{B_{v}}\sum_{b=1}^{B_{v}}\left\{ \left[\frac{-1}{2\sigma_{\delta,m}^{\left(t\right)2}}+\frac{\left(\delta_{m}^{\left(b\right)}-\mu_{\delta,m}^{\left(t\right)}\right)^{2}}{2\sigma_{\delta,m}^{\left(t\right)4}}\right]\cdot g_{\boldsymbol{\eta}}^{\left(t,b\right)}\right\} .
\end{align}
The derivation is similar for $\sigma_{\phi,m}^{2}$.

\bibliographystyle{IEEEtran}
\bibliography{MM-SPVBI}

\end{document}